\documentclass[
  aps, prd, reprint, superscriptaddress,
  amsmath, amssymb, nofootinbib, twocolumn,
  floatfix
]{revtex4-2}
\usepackage{xcolor}
\usepackage{graphicx}% Include figure files
\usepackage{dcolumn}% Align table columns on decimal point
\usepackage{hyperref}% add hypertext capabilities
\usepackage[caption=false]{subfig}
\usepackage[normalem]{ulem} % useful for editing, e.g. strikethrough, etc.
\usepackage{comment} % to comment out large portions of the text

\newcommand{\sk}[1]{\textcolor{black}{#1}}

\newcommand{\priorname}{\textsc{Goldilocks prior}}
\newcommand{\lt}{\tilde{\Lambda}}
\newcommand{\msol}{\text{M}_{\odot}}
\newcommand{\dlt}{\delta\tilde{\Lambda}}
\newcommand{\mc}{\mathcal{M}_c}
\newcommand{\secname}{Section}
\newcommand{\tabname}{Table}
\newcommand{\figname}{Figure}
\newcommand{\sublabel}[2]{%
\protected@edef\@currentlabel{#1(#2)}%
  \label{#1#2}%
}

\usepackage{newtxtext}
\usepackage{newtxmath}
\usepackage{bm}% bold math
\begin{document}
\title{The Goldilocks Prior: Physically Consistent Inference of Neutron Star Properties from Gravitational Wave Observations}
\raggedbottom
\author{Sanika Khadkikar}
\affiliation{Institute for Gravitation and the Cosmos, Department of Physics, Pennsylvania State University, University Park, PA 16802, USA}
\affiliation{Department of Physics, The Pennsylvania State University, University Park, PA 16802, USA}

\author{Bangalore Sathyaprakash}
\affiliation{Institute for Gravitation and the Cosmos, Department of Physics, Pennsylvania State University, University Park, PA 16802, USA}
\affiliation{Department of Physics, The Pennsylvania State University, University Park, PA 16802, USA}
\affiliation{Department of Astronomy \& Astrophysics, The Pennsylvania State University, University Park, PA 16802, USA}

\begin{abstract}
Gravitational wave observations of neutron star mergers provide a unique window into the interiors of neutron stars, probing dense matter from microscopic nuclear interactions to the macroscopic structure, including their maximum mass, beyond which the remnant collapses to a black hole. Accurate inference of these properties is crucial, as many astrophysical, nuclear, and multimessenger studies rely on gravitational wave constraints. Studies on estimating neutron-star properties from gravitational waves often employ priors that neglect correlations between parameters, or treat tidal degrees of freedom inconsistently, leading to biases in inferred neutron-star properties. We introduce a novel physically informed prior, \priorname, that is informed by a broad, semi-parametric ensemble of equations of state, accounts for the joint correlations among the chirp mass, mass ratio, and both tidal deformability parameters, and is tailored to detector sensitivity. Using several mock binary neutron star merger simulations spanning symmetric and asymmetric masses, spinning and non-spinning systems, and a range of signal-to-noise ratios, \sk{we show that \priorname\ consistently reduces biases in the inferred tidal posteriors relative to uniform and previously proposed EoS-informed priors}. This framework provides a robust, systematic approach for neutron star parameter estimation, enabling more reliable inference for both current and future gravitational wave observations.

\end{abstract}

\maketitle

\section{Introduction}
\label{sec:intro}
Neutron stars (NSs) are among the most extreme objects in the universe, with densities several times larger than the nuclear saturation density \cite{Lattimer:2012nd, Lattimer:2015nhk, Oertel:2016bki, Paschalidis:2016vmz, Ozel:2016oaf, Baym:2017whm}. Unlike black holes, NSs emit information across multiple observational channels, providing a unique window into high-energy physics and astrophysics. The dense matter equation of state (EoS) plays a central role in the dynamics of NSs, as it determines the relationship between the pressure and density and thereby impacts macroscopic observables such as the NS mass and radius \cite{Prakash:1996xs, Lattimer:2015nhk}. Determining the EoS remains one of the most important open problems in modern astrophysics, as it provides insight into the microphysical behaviour of matter under extreme conditions and also informs our understanding of the formation and evolution of NSs.

Gravitational wave (GW) observations provide a direct channel for studying NSs and constraining their EoS \cite{Schutz:1999xj, Read:2009yp}. The LIGO-Virgo-KAGRA (LVK) \cite{LIGOScientific:2014pky, VIRGO:2014yos, KAGRA:2020tym} network of GW detectors is sensitive to compact object mergers involving NSs, having already reported a few candidate events consistent with binary neutron star (BNS) and neutron star black hole (NSBH) mergers \cite{LIGOScientific:2018mvr, LIGOScientific:2020ibl, KAGRA:2021vkt, LIGOScientific:2025slb}. Several fundamental NS properties, such as masses ($m$), spins ($\chi$), and tidal deformabilities ($\Lambda$), can be inferred by analysing the observed GW amplitude and phase evolution \cite{Farr:2015lna, Chatziioannou:2020pqz}. The phase evolution of a GW signal from a BNS or NSBH system, at leading order in the post-Newtonian approximation, is dominated by its chirp mass $\mc$ and the mass ratio $q$ \citep{Blanchet:2013haa}. Higher-order post-Newtonian terms contain spin effects, and the tidal effects first enter at the fifth post-Newtonian order (i.e., at ${\cal O}(v/c)^{10}$), where the signal is primarily sensitive to the \emph{effective tidal deformability,} $\lt$. The next-to-leading order tidal term depends on a quantity called the \emph{reduced tidal deformability,} $\dlt$ \cite{Flanagan:2007ix, Hinderer:2007mb, Favata:2013rwa}.  While alternative parameterizations are possible, we adopt the leading-order combination $(\mc, q, \lt, \dlt)$ throughout this work for convenience. Any such equivalent set of parameters can then be used to constrain the NS EoS, directly linking GW observations to the physics of dense matter.

Inference of NS source parameters from GWs typically relies on Bayesian inference techniques. These analyses often employ broad, ``uninformative"\footnote{Any prior with compact support is not uninformative in a strict Bayesian sense. We use this terminology to imply that the prior in the relevant parameter is assumed to be flat.} priors to ensure that the resulting posterior distributions are driven primarily by the data rather than by predefined assumptions. In practice, this usually corresponds to uniform distributions over carefully chosen, broad parameter ranges intended to accommodate a wide variety of possible NS structures (see Ref.\,\cite{LIGOScientific:2018hze} for an example implementation). 

This approach is perfectly reasonable when GW signals have sufficiently high signal-to-noise ratios (SNRs), as the information content of the data dominates over any previously assumed priors. For next-generation detectors such as the US-based Cosmic Explorer \cite{Reitze:2019iox, Evans:2021gyd, LIGOScientific:2016wof}, and the European Einstein Telescope \cite{Punturo:2010zza, Hild:2010id, ET:2019dnz}, substantially improved sensitivity and bandwidth are expected to yield a much higher rate of high-SNR detections than current-generation instruments \cite{Gupta:2023lga}. However, such loud signals from BNS or NSBH systems are expected to be relatively rare given the sensitivities of currently operating GW detectors and the planned LVK upgrades over the next decade (see \cite{LIGOScientific:2025slb}, \cite{Gupta:2023lga}, and Figure 3 in \cite{Khadkikar:2025ith}). Thus, in the near future, for GW detections involving NSs prior assumptions will substantially impact the inferred source properties.

As a consequence, analyses of lower-SNR signals often rely on broad uniform priors to allow the data to contribute wherever information is present in the GW signal. However, this reliance can become problematic when physically plausible regions of the NS parameter space receive little to no support from the priors. As an example, uniform priors on commonly used tidal parameter combinations can effectively undersample low tidal deformability values, resulting in a systematic bias toward higher inferred tidal deformabilities as detailed in \secname\ \ref{sec:LVK_priors}. 

Moreover, while allowing for a wide range of possible NS physics is important, NS parameters such as $m$ and $\Lambda$ exhibit strong, physically motivated correlations \cite{Flanagan:2007ix, Hinderer:2007mb, Hinderer:2009ca}. Including these correlations is important because they reflect the NS’s fundamental response to its companion’s tidal field during a merger. Adopting independent uniform priors for such parameters that are naturally correlated can lead to exploration of regions of parameter space that are inconsistent with known physics, potentially biasing the inferred constraints. Recognising both of these limitations motivates the development of approaches that more faithfully capture the underlying physics, which we present in this work.

Some previous studies have recognised and discussed the consequences of prior choice in GW analyses and outlined methods to mitigate its effects. One widely used approach is reweighting, which allows posterior distributions obtained with one prior to be statistically transformed to reflect an alternative prior at a fraction of the computational cost of resampling (see, e.g., \cite{Biscoveanu:2021cpd, Chattopadhyay:2024hsf}). Reweighting can correct for the influence of a chosen prior when the relevant regions of parameter space are adequately sampled. However, if portions of the parameter space are undersampled or physically valid regions are not explored, as discussed above with independent uniform priors, reweighting cannot fully compensate \cite{Prasad:2026idz}. This reinforces the importance of adopting physically informed priors from the outset to ensure that posterior inferences remain consistent with known NS physics.

All of the aforementioned effects may seem subtle, but they can induce systematic biases in the NS properties inferred from GWs. To prevent such biases from propagating, the priors used for inferring NS parameters should be informed by the expected behaviour of physically plausible EoSs. A wide range of parametric EoS models exist, including piecewise polytropes \cite{Read:2009yp}, spectral representations \cite{Lindblom:2010bb}, meta-models \cite{Margueron:2017eqc}, and sound-speed parametrisations \cite{Tews:2018kmu, Greif:2018njt}. Recently, non-parametric representations based on Gaussian processes have also been introduced to generate causal and thermodynamically stable EoSs without specifying an explicit functional form \cite{Landry:2018prl, Gorda:2022jvk, Mroczek:2023eff}. 

For this study, we use the large ensemble of physically plausible EoSs made publicly available by \citet{Ng:2025wdj} to inform our tidal priors. These EoSs are constructed using a semi-parametric approach that combines meta-modelling constrained by nuclear theory and experiments at low density with Gaussian-process-generated extensions at higher densities. By drawing from this ensemble, we can construct priors on tidal deformabilities that are dependent on the corresponding NS masses, naturally capturing the expected correlations between $m$ and $\Lambda$.

While significant progress has been made in constructing ensembles of possible NS EoSs, it is not guaranteed that any chosen ensemble fully spans the true NS EoS. More importantly, biases can arise even when the true EoS lies within the ensemble. Detector noise can perturb the phase and amplitude of GW signals enough to shift the inferred constraints on NS properties away from their true values. If the prior is tightly constrained by a limited EoS ensemble, it may restrict the parameter space more strongly than the measurement uncertainty warrants. In this case, the posterior can become artificially concentrated near the boundaries imposed by the prior, leading to biased inferences. Therefore, priors should be constructed to reflect both the range of tidal deformabilities predicted by the EoS ensemble and the expected measurement uncertainty of the GW detector network, allowing the inference to remain driven primarily by the data rather than by prior boundaries.

Previous efforts, such as \citet{Magnall:2025zhm}, have explored two-dimensional joint priors linking $\mc$ and $\lt$, demonstrating that incorporating correlations can alter posterior inferences. However, such approaches are restricted to two dimensions and do not capture the full multidimensional structure of the tidal parameter space. In particular, $\dlt$ was assigned an independent uniform prior in Ref.\,\cite{Magnall:2025zhm}, motivated by its limited measurability at current detector sensitivity levels \cite{Wade:2014vqa}. While $\dlt$ is indeed weakly constrained by present GW observations, it nevertheless occupies a restricted and physically well-defined range and exhibits strong correlations with other tidal parameters.

More recently, \citet{Wouters:2025csq} constructed a data-inspired neural prior that captures the correlations of the tidal parameters with the masses, learning their joint distribution from a phenomenological EoS parameterisation. While such parametric EoS representations provide a useful and computationally tractable framework, they may not fully capture the breadth of physically plausible nuclear matter behaviour, particularly in regimes where theoretical uncertainties remain large. In particular, some parameterisations can underrepresent the diversity of high-density compositions, where nucleonic, hyperonic, and quark matter phases may contribute in ways that are difficult to encode within a single functional form.

By contrast, realistic analyses can benefit from semi-parametric EoS ensembles that are strongly constrained at low densities by crust physics and meta-model descriptions, while retaining greater flexibility at higher densities where observational and theoretical guidance is weaker. Such ensembles are designed to span a broader class of physically motivated models, rather than being confined to a single phenomenological parameterisation, thereby offering a more comprehensive representation of current EoS uncertainty.

Both \citet{Magnall:2025zhm} and \citet{Wouters:2025csq} represent meaningful progress toward the shared goal of improving NS prior inference in GW parameter estimation. Building on these developments, this work presents a prior framework that aims to overcome several limitations of previous approaches. We construct a novel prior over the full tidal parameter space, informed by a semi-parametric EoS ensemble and the expected detector response, thereby ensuring physically motivated prior support. Since our goal is to develop a prior that is neither overly informative nor purely agnostic, we refer to this conditional prior as the \priorname.

We do this by assuming a uniform distribution of NS masses and constructing a four-dimensional conditional prior in the ($\mc$, $q$, $\lt$, $\dlt$) space. The choice of mass population distribution is flexible and can be replaced by alternative distributions according to one's preference. Similarly, the specific choice of the two tidal dimensions does not significantly affect the results, as discussed later in this work.

We then perform a detailed Fisher analysis to estimate the median uncertainty in $\tilde{\Lambda}$ as a function of its true value for the considered BNS systems with the current-generation LVK network sensitivity. Combining the EoS ensemble with the expected detector uncertainty, we construct the \priorname\footnote{The implementation, an example prior grid and a runnable \texttt{bilby} example are available at \url{https://github.com/sanikakhadkikar/Goldilocks-Prior}.}.

Using this new prior, we perform extensive Bayesian parameter inference on mock data of BNS systems, with a range of tidal deformabilities and located at varying distances, resulting in different SNRs. We use \textsc{IMRPhenomXAS\_NRTidalv3} for injection and recovery because its calibration includes an expanded suite of numerical relativity data covering high-mass ratios, a broader range of EoSs, and dynamical tidal effects \cite{Pratten:2020fqn,  Abac:2023ujg}. Across the mock GW simulations studied, the Goldilocks prior yields the smallest prior-driven bias in the $\lt$ and $\dlt$ posteriors relative to uniform priors and the prior of \citet{Magnall:2025zhm}. Given the broad scope of this work, a follow-up study applying this approach to NSBH systems and real GW data will be presented in companion publications.

The rest of this paper is organised as follows. \secname\ \ref{sec:LVK_priors} describes the standard priors used for analysing GWs from BNS mergers, with \secname\ \ref{subsec:support} highlighting regions of $\lt$ space with minimal or no support and \secname\ \ref{subsec:correlations} identifying prior volume with physically inconsistent regions in the NS parameter space. \secname\ \ref{sec:newprior} presents the construction of the new prior, including \secname\ \ref{subsec:eos}, which details the EoS ensemble, \secname\ \ref{subsec:sensitivity}, which describes the incorporation of detector-dependent $\lt$ uncertainties, and \secname\ \ref{subsec:multid}, which explains the creation of a multidimensional conditional prior. We present the results of our study in \secname\ \ref{sec:results} and conclude in \secname\ \ref{sec:conclusions} with a summary of our findings and future prospects.

\section{State of the art}

\begin{figure*}[t]
    \centering
    \includegraphics[width=\linewidth]{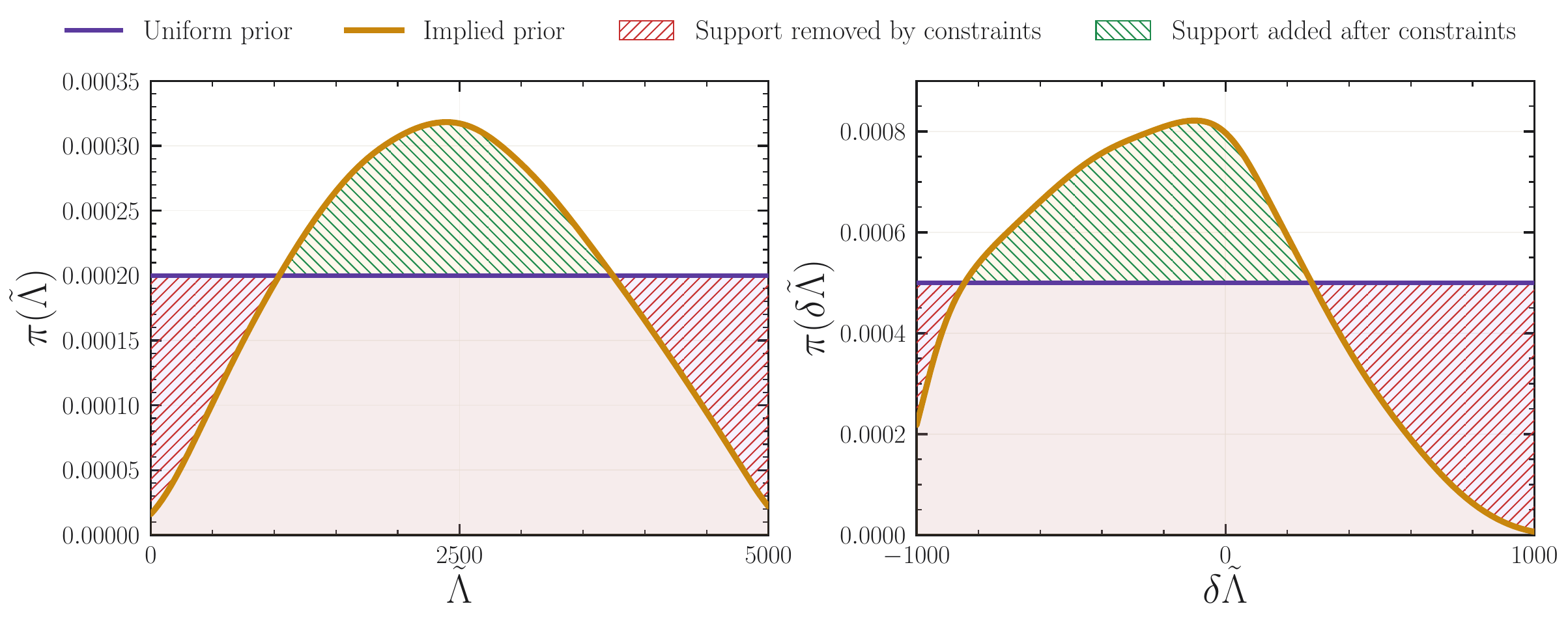}
    \caption{\sk{Implied prior distributions of $\lt$ (\emph{left}) and $\dlt$ (\emph{right}). Assuming uniform priors on $\lt \in U[0,5000]$ and $\dlt \in U[-1000, 1000]$, together with constraints on $\Lambda_{1,2} \geq 0$,  produces this emergent prior structure from 2 million samples. The flat purple line is the true uniform prior, and the orange curve is the implied prior after the constraint. Green hatching marks where the constraint adds support relative to the uniform prior, while red hatching indicates where it removes support. The constraint suppresses support for both high and low $\lt$ values, while $\dlt$ is pushed toward zero, with its negative range disproportionately favoured.}}
    \label{fig:1_lt_support}
\end{figure*}
\label{sec:LVK_priors}
GWs encode information about several source properties of NSs, including their masses ($m_i$), spins ($\chi_i$), and tidal deformabilities ($\Lambda_i$), where $i = 1,2,$ labels the two NSs in the binary. In a Newtonian framework, the tidal deformability of a component is related to its mass via
\begin{equation}
\label{eq:lambda}
\Lambda_i = \frac{2}{3} k_2 \left( \frac{R c^2}{G m_i} \right)^5,
\end{equation}
where $R$ denotes the NS radius, $k_2$ the quadrupolar tidal Love number, $c$ the speed of light, and $G$ the gravitational constant \cite{Flanagan:2007ix}. The relationship between $\Lambda$ and  $m$ is determined by the values of $k_2$ and $R$, which are in turn set by the NS EoS. Although $m_i$ and $\Lambda_i$ can be inferred from GW observations, GW parameter estimation typically employs a parameterisation that follows the leading-order mass and tidal contributions in the post-Newtonian expansion. Consequently, GW analyses use parameter combinations that naturally arise within the post-Newtonian formalism. Rather than  $m_i$, the chirp mass, $\mathcal{M}_c = \frac{(m_1 m_2)^{3/5}}{M^{1/5}}$, where $M = m_1 + m_2$ is the total mass, is used together with a second mass parameter. This second parameter is commonly taken to be either the symmetric mass ratio, $\eta = \frac{m_1 m_2}{M^2}$, or the mass ratio, $q = \frac{m_2}{m_1} \leq 1$. In this work, we adopt $q$ as the second mass parameter.

Similarly, tidal effects enter the waveform primarily through the leading-order 
\begin{equation}
\label{eq:lt}
\tilde{\Lambda} = \frac{16}{13} \frac{(m_1 + 12 m_2)m_1^4 \Lambda_1 + (m_2 + 12 m_1)m_2^4 \Lambda_2}{(m_1 + m_2)^5},
\end{equation}
and the next-to-leading-order contribution is captured by 
\begin{align}
\delta \tilde{\Lambda} &= \frac{1}{2} \Bigg[ 
\sqrt{1 - 4\eta} \left( 1 - \frac{13272}{1319} \eta + \frac{8944}{1319} \eta^2 \right)(\Lambda_1 + \Lambda_2) \nonumber \\
& \quad + \left( 1 - \frac{15910}{1319} \eta + \frac{32850}{1319} \eta^2 + \frac{3380}{1319} \eta^3 \right)(\Lambda_1 - \Lambda_2) 
\Bigg]
\label{eq:3_dlt}
\end{align}
Thus, the ($\lt, \dlt$) combination can also be used as opposed to the ($\Lambda_1, \Lambda_2$) combination. Throughout this work, we adopt $\mathcal{M}_c$ and $q$ as the mass parameters, and $\tilde{\Lambda}$ and $\delta \tilde{\Lambda}$ as the tidal parameters.

Generally, Bayesian parameter estimation techniques are used to extract these NS source properties from the GW data observed. Let the set of intrinsic source parameters, together with the extrinsic parameters required to model an observed GW signal, be denoted by $\vec{\theta}$, and let $d$ represent the observed GW data. Bayes’ theorem then yields the posterior probability distribution for the parameters $\vec\theta$ given the data,
\begin{equation}
   p(\vec{\theta} \mid d) = \frac{\mathcal{L}(d \mid \vec{\theta})\,\pi(\vec{\theta})}{Z(d)},
\end{equation}
where $\mathcal{L}(d \mid \vec{\theta})$ is the likelihood, $\pi(\vec{\theta})$ is the prior distribution, and $Z(d)$ is the normalisation constant, commonly referred to as the evidence.

As discussed in \secname\ \ref{sec:intro}, for the current generation of LVK detectors the joint prior distribution, $\pi(\vec{\theta})$, can play a dominant role relative to the likelihood, $\mathcal{L}(d \mid \vec{\theta})$, motivating a careful examination of its construction. For the extrinsic parameters in $\vec{\theta}$, it is standard to assume that binaries are distributed isotropically in sky position and orientation \cite{LIGOScientific:2025slb}. The primary modelling choices, therefore, concern the intrinsic parameters. 

NS masses are typically assigned uniform priors on the component masses, from which $\mc$ and $q$ are derived. Since $\mc$ is tightly constrained by matched-filter searches, the maximum and minimum bounds placed on $\mc$ can be correspondingly narrow \citep{Ewing:2023qqe}. The specific choices adopted in this work are described in \secname\ \ref{sec:results}. Spin priors are informed by observations of Galactic BNSs, which indicate that even the fastest-spinning systems expected to merge within a Hubble time, such as PSR J0737–3039A \cite{Burgay:2003jj} and PSR J1946+2052 \cite{Stovall:2018ouw}, will have dimensionless spins of at most $\chi \sim 0.04$–$0.05$ at merger. Motivated by this, GW analyses generally adopt low-spin priors with $\chi_i \leq 0.05$ \cite{LIGOScientific:2018mvr, LIGOScientific:2018hze}. For completeness, alternative high-spin priors ($\chi_i \leq 0.89$) are also commonly considered, and parameter estimation is often performed under multiple spin assumptions \cite{LIGOScientific:2018hze}.

Given an EoS, $\Lambda_i$ of each star is uniquely determined by $m_i$. Consequently, as seen from Eqs.~(\ref{eq:lambda})--(\ref{eq:3_dlt}), $\lt$ and $\dlt$ are likewise completely determined by the component masses alone and do not constitute independent degrees of freedom given the EoS. In such an idealised scenario, posteriors for all tidal parameters could, in principle, be obtained directly from the inferred mass posteriors. In practice, however, this is not possible because the NS EoS remains largely unknown.

As a consequence of this uncertainty, we would need to treat the tidal deformabilities as independent parameters. Tidal priors remain poorly constrained, owing both to the limited number of astrophysical observations and to the wide range of physically plausible values predicted by existing EoS models. Millions of candidate EoSs in the literature predict markedly different maximum values for $\Lambda$ over the relevant mass range, motivating the common choice of a conservative upper bound $\Lambda_i^{\rm max}=5000$\footnote{Since the constraint $m_1\ge m_2$ is typically imposed in GW analyses, $\Lambda_2^{\rm max}$ is sometimes extended to 10000 to account for this ordering.}. The formalism of \citet{Flanagan:2007ix} further requires $\Lambda$ to be non-negative for NSs, setting the lower bound $\Lambda_i^{\rm min} = 0$. Given the small number of BNS detections to date, the data provide little guidance on the shape of the prior, and a uniform distribution between these bounds is therefore commonly adopted as a broad and minimally informative choice \cite{LIGOScientific:2018hze, LIGOScientific:2020aai}. Analogous uniform priors are often employed for the reparameterised tidal quantities, with $\lt \in U[0,5000]$ and $\dlt \in U[-1000,1000]$, where $\lt$ is typically the more tightly constrained combination. Here, $U$ denotes a uniform distribution.

It is important to note that choosing uniform priors in either the component tidal deformabilities, $(\Lambda_1, \Lambda_2)$, or in the combinations $(\tilde{\Lambda}, \delta\tilde{\Lambda})$ leads to qualitatively similar results. For priors uniform in $(\Lambda_1, \Lambda_2)$, the parameters must ultimately be converted to $\tilde{\Lambda}$ for waveform evaluation, since it is the leading-order tidal parameter. Conversely, when using uniform priors in $(\lt, \dlt)$, the physical constraint that $\Lambda_{1,2} \geq 0$ must also be enforced. In both cases, these transformations and constraints produce roughly equivalent prior support, so the choice of parameterisation does not qualitatively change the analysis. We adopt the $(\lt, \dlt)$ combination for this study.
\subsection{Regions of limited support in the parameter space}
\label{subsec:support}

\begin{figure*}
    \centering
    \includegraphics[width=\linewidth]{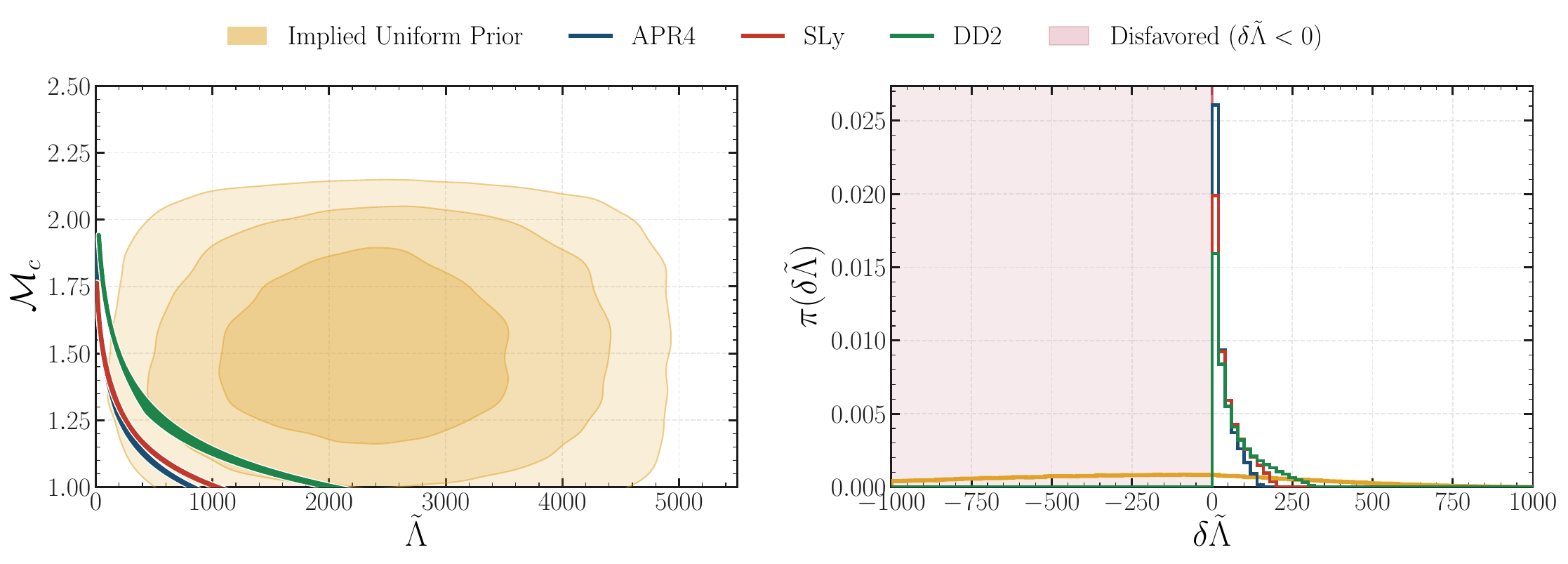}
    \caption{Comparison between the standard prior described in \secname\ \ref{sec:LVK_priors} and the tidal deformability ranges predicted by physically motivated EoSs consistent with existing observational constraints. The \textsc{APR4} EoS (\emph{blue}) predicts smaller $\lt$ values, \textsc{SLy} (\emph{red}) predicts intermediate values, and \textsc{DD2} (\emph{green}) predicts larger values. The left panel shows the two-dimensional prior in the $\mc$–$\lt$ plane, while the right panel shows the corresponding prior on $\dlt$. \sk{The shaded red band marks the physically disfavoured $\dlt <0$ region}. The standard uniform prior spans a substantially larger region of the $\mc$–$\lt$ space than is supported by the EoS bounds. More importantly, the implied $\dlt$ prior from the standard choice differs markedly from that obtained when physically plausible EoSs are imposed.}
    \label{fig:2_mc_lt_dlt}
\end{figure*}

For the chosen parameterisation $(\lt, \dlt)$, standard analyses typically adopt uniform priors together with the constraint $\Lambda_{1,2} \geq 0$, as described above. These constraints induce a non-uniform \emph{implied} prior in $(\lt, \dlt)$, shown in \figname\ \ref{fig:1_lt_support}. In addition to the tidal parameters, we sample uniformly in the component masses and use these to transform consistently between the two tidal parameterisations. The implied prior on $\lt$ in the left panel in \figname\ \ref{fig:1_lt_support} peaks around $\lt \approx 2500$ and has very low support at both the smallest and largest values of $\lt$. The lack of support at low $\lt$ arises because only a \sk{single} combination of $\Lambda_1$ and $\Lambda_2$, namely, $\Lambda_1 = \Lambda_2 = 0$, produces $\lt =0$. By the same logic, lower $\lt$ values ($\lesssim 100$) correspond to relatively few combinations of $\Lambda_{1,2}$, and therefore receive comparatively low prior support.

Similarly, the right panel shows that $\dlt$ develops a non-trivial structure due to the same constraints: support is concentrated near $\dlt = 0$, biased toward negative values, and progressively decreases for positive $\dlt$ values. Together, these implied prior shapes demonstrate that even a supposedly uninformative uniform prior is already informative, indirectly favouring higher $\lt$ and negative $\dlt$ values. The structure of both $\lt$ and $\dlt$ priors \sk{indicates} a systematic overestimation in inferred tidal parameters that can affect GW analyses if not properly accounted for.

As expected from these emergent priors, \citet{Khadkikar:2025ith} observed a systematic overestimation of inferred $\lt$ for EoSs such as \textsc{APR4} (\cite{Akmal:1998cf}) and \textsc{SLy} (\cite{Douchin:2001sv}), which predict lower and intermediate $\lt$ values. The study considered not only the current generation of GW detectors but also planned LVK upgrades and next-generation detector networks, including Cosmic Explorer and the Einstein Telescope. Similar overestimation trends were observed for the next LVK upgrades to the $A^{\#}$ sensitivity for both \textsc{APR4} and \textsc{SLy} EoSs \cite{T2200287}. While the bias was significantly reduced for highly sensitive next-generation GW detectors, heavier mass systems having low tidal deformabilities (see Eq. \ref{eq:lambda}) continued to exhibit this overestimation.

These results highlight that systematic biases induced by such implied tidal priors directly impact NS property inference and, consequently, constraints on the NS EoS. Even with highly precise next-generation detectors, heavier BNS systems that are important for determining the mass at which NSs collapse to black holes remain susceptible to these biases. It is therefore essential to adopt physically motivated priors that both reflect the full plausible parameter space and have well-defined distributions to avoid biasing tidal parameter inferences.

\subsection{Physically inconsistent regions in the parameter space}
\label{subsec:correlations}

According to Eq. \ref{eq:lambda}, heavier NSs are expected to have lower $\Lambda$ values, while lighter NSs have larger $\Lambda$ values. Incorporating this physically motivated relation into the prior already provides a clear advantage, as it reduces the effective prior volume and, consequently, the sampling cost of parameter estimation analyses. This effect is illustrated in the left panel of \figname\ \ref{fig:2_mc_lt_dlt}, which shows the two-dimensional distribution in the $\mc$-$\lt$ plane.

The prior distributions in \figname\ \ref{fig:2_mc_lt_dlt} are generated using $10^6$ samples drawn from mock mass distributions that are uniform in the component masses, $m_{1,2} \in U[1,2.5]$, with the ordering $m_1 \geq m_2$, along with the uniform tidal priors as described in \secname\ \ref{subsec:support}. The left panel compares the region populated by the standard uniform prior with the region spanned by representative EoSs that are consistent with current GW observations \cite{LIGOScientific:2018cki}. In particular, \textsc{APR4} EoS shown in blue predicts smaller $\lt$ values, \textsc{SLy} in red predicts intermediate values, and \textsc{DD2} in green predicts larger, more extreme $\lt$ values. It is evident that the uniform prior covers a substantially larger region of the $\mc$-$\lt$ space than that occupied by these chosen, representative set of physically plausible EoSs.

The right panel highlights the significant impact also on $\dlt$. Here, the implied $\dlt$ prior from \figname\ \ref{fig:1_lt_support} is compared with priors constructed from the same three EoSs. The EoS-based prior, denoted $\pi(\vec{\theta})_{\mathrm{EoS}}$, is obtained by drawing samples uniform in component masses as described above and mapping them to tidal deformabilities using the corresponding EoS predictions. In contrast to the broad prior induced by uniform assumptions, the EoS-based representative conditional priors yield a much narrower $\dlt$ distribution, sharply centred around $\dlt = 0$. Moreover, whereas the uniform prior \sk{places a substantial weight on negative $\dlt$ values}, the EoS-based prior shows almost no support for negative $\dlt$ values and restricts the positive range to a much narrower interval. 

This comparison shows that, even though $\dlt$ is weakly constrained by current GW data, its physically allowed range is far more restricted than is assumed under standard priors. A narrow, physically motivated prior on $\dlt$ therefore has the potential to significantly improve inference, rather than being safely ignored. This provides a clear motivation for adopting EoS-informed priors throughout this work, as they naturally encode the relevant mass–tidal correlations and avoid unnecessary exploration of unphysical regions of parameter space.

\subsection{\sk{Post-processing to change priors}}
\label{subsec:reweighting}

\begin{figure*}[b]
    \centering
    \includegraphics[scale=0.5]{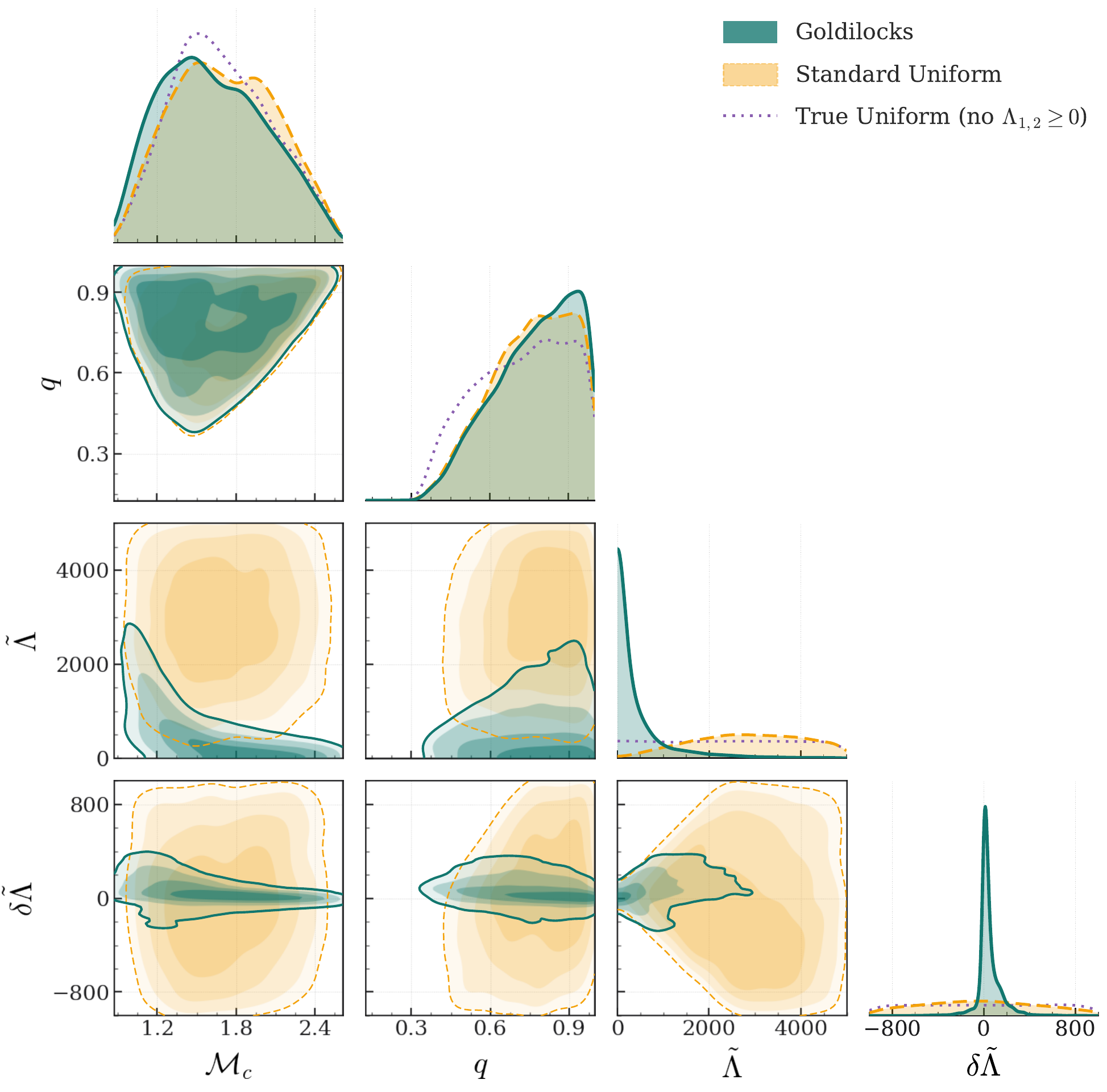}
    \caption{A comparison between the new \priorname\ proposed in this work (teal) and the standard uniform priors commonly used in the literature (orange), shown as a corner plot. \sk{The dotted curve shows a true uniform prior without the $\Lambda_{1,2} \geq 0$ constraint applied}. Two million samples are generated for each prior, with contours from one sigma to four sigma. The $\mc$–$\lt$ projection demonstrates, consistent with the discussion in \secname\ \ref{subsec:correlations}, that physically motivated equations of state substantially reduce the allowed prior volume in this plane. This correlation leaves an imprint on the correlations of the tidal parameters with the mass ratio $q$ as well. The projections involving $\dlt$ further show that this parameter is significantly more constrained under the new prior, in contrast to the broader distributions produced by the standard uniform priors.}
    \label{fig:3_prior_comparison}
\end{figure*}
The distortions described above do not, by themselves, force a re-analysis. If a prior is judged unsuitable after the Bayesian inference is carried out, the posterior it produced can be mapped onto a different prior without repeating the computationally expensive sampling. Given posterior samples drawn under $\pi_1(\vec{\theta})$, the posterior corresponding to an alternative prior $\pi_2(\vec{\theta})$ follows from
\begin{equation}
    p_2(\vec{\theta} \mid d) \propto p_1(\vec{\theta} \mid d) \, \frac{\pi_2(\vec{\theta})}{\pi_1(\vec{\theta})},
    \label{eq:reweighting}
\end{equation}
which requires only that $\pi_2$ have support wherever $\pi_1$ does. Since a full parameter estimation run over this many dimensions is expensive, this is an attractive alternative to sampling afresh, and it is widely used \cite{Biscoveanu:2021cpd, Chattopadhyay:2024hsf}.

The condition on shared support is, strictly, satisfied. The implied uniform prior does not vanish at $\lt = 0$, since the single combination $\Lambda_1 = \Lambda_2 = 0$ produces it, so $\pi_1$ remains nonzero and the ratio $\pi_2/\pi_1$ in Eq.~\ref{eq:reweighting} stays finite everywhere. The difficulty is practical rather than formal. Reweighting is applied not to densities but to a finite set of samples, each assigned a weight $w_i = \pi_2(\vec{\theta}_i)/\pi_1(\vec{\theta}_i)$. The reweighted posterior is assembled from whichever samples the original run happened to draw, rescaled, and no new samples are produced. Since $\pi_1$ places very little support at low $\lt$, as shown in \figname~\ref{fig:1_lt_support}, few samples are drawn there, and those few receive the largest weights. The recovered density in this region is then built from a handful of points, producing spurious spikes rather than a faithful reconstruction. We quantify this in \secname~\ref{sec:results} and Appendix~\ref{app:reweighting}.
\section{A balanced, physics-informed prior}
\label{sec:newprior}

A prior that is neither blind to the physically motivated relationship between $\mc$ and $\lt$ dictated by the dense matter EoS, nor overly constrained by any specific EoS realisation, provides a natural improvement over the shortcomings of agnostic uniform priors discussed above. In this work, we propose a three-part framework to construct such a prior, designed to reduce biases and faithfully capture physically allowed correlations. First, the prior is informed by a broad ensemble of physically plausible EoSs to capture the expected correlations between $\mc, q, \lt$ and $\dlt$ as highlighted in \secname\  \ref{subsec:eos}. Second, to account for the possibility that a noise realisation at the time of detection might affect $\lt$ and $\dlt$ inference, additional flexibility is introduced through detector sensitivity, as described in \secname\ \ref{subsec:sensitivity}. This ensures the prior remains robust to uncertainties ushered in by noise while incorporating realistic measurement constraints. Third, the construction explicitly includes $\dlt$ alongside $\mc, q$ and $\lt$, producing a fully four-dimensional conditional prior that consistently captures all tidal correlations and eliminates biases arising from the inconsistent treatment of $\dlt$, as detailed in \secname\  \ref{subsec:multid}.

\subsection{Neutron Star Physics}
\label{subsec:eos}
We use $105{,}000$ EoSs from the \citet{Ng:2025wdj} dataset to construct the ensemble that informs our priors. For each EoS, we generate a mass grid of $1000$ $(m_1, m_2)$ samples and compute the corresponding $\Lambda_{1,2}$ values. We assume a uniform distribution in both masses, though this choice can be modified as desired. We adopt $m_{\rm min} = 1\ \text{M}_{\odot}$ as the minimum mass and set the maximum mass for each EoS to $m_{\rm max}^{\rm EoS}$, the maximum cold NS mass allowed by that EoS. EoSs with $m_{\rm max}^{\rm EoS} < 2~\text{M}_{\odot}$ are rejected, consistent with observations of the heaviest known pulsars, whose median mass values exceed $2~\text{M}_{\odot}$ \cite{Antoniadis:2013pzd, NANOGrav:2019jur, Fonseca:2021wxt}. However, it is important to note that the threshold values $\{m_{\rm min}^{\rm EoS}, m_{\rm max}^{\rm EoS}\}$ can be changed as needed. The resulting $(m_1, m_2, \Lambda_1, \Lambda_2)$ samples are then converted to $(\mc, q, \lt, \dlt)$ for subsequent analyses.

An explicit comparison between the priors described in \secname\ \ref{sec:LVK_priors} and the \priorname\ is shown in \figname\ \ref{fig:3_prior_comparison}. This figure clearly illustrates the limitations of uniform priors highlighted in \secname\ \ref{subsec:correlations}. In this corner plot, the distributions of two million $\mc$, $q$, $\lt$ and $\dlt$ samples are shown for both priors. For the $\lt$ distribution, the \priorname\ depicted in teal is significantly narrower than the uniform prior in orange, reflecting the physical constraints imposed by realistic EoSs. The expected inverse relationship between $\Lambda$ and $m$ is also visible in the $\lt$-$\mc$ plot. \sk{While the $\mc$-$q$ relation is standard, the EoS ensemble additionally couples $q$ to the tidal parameters, further reducing the prior volume}. We also discuss the origin of the power-law tail of the one-dimensional prior distribution obtained for $\tilde{\Lambda}$ in Appendix~\ref{app2}.

The bottom row shows that $\dlt$ is more tightly constrained under the \priorname, with substantial support concentrated near $\dlt = 0$. We impose the mass ordering constraint $m_1 \geq m_2$  for drawing component mass samples, as is commonly done. The constraint produces a positive skew in $\dlt$, \sk{as already visible in the EoS-based examples of \figname\ \ref{fig:2_mc_lt_dlt}}. The ordering generates a large fraction of nearly symmetric systems, which naturally yield $\dlt = 0$. For increasingly asymmetric systems, the same constraint biases $m_1$ toward larger values and $m_2$ toward smaller values. Combined with the inverse mass–tidal deformability relation, this favours lower $\Lambda_1$ and higher $\Lambda_2$, resulting in $\dlt \geq 0$ when evaluated using Eq.~\ref{eq:3_dlt}.

\sk{In the \priorname\ of \figname\ \ref{fig:3_prior_comparison}, this positive skew and the narrow $\mc$-$\lt$ correlation are less sharply defined compared to the noise-free EoS priors of \figname\ \ref{fig:2_mc_lt_dlt}. This broadening is introduced deliberately through the detector-sensitivity noise described in \secname\ \ref{subsec:sensitivity}. The reductions in prior volume, particularly in $\dlt$, significantly aid in constraining the posteriors, as demonstrated in \secname\ \ref{sec:results}.}

\subsection{Detector Sensitivity Dependence}
\label{subsec:sensitivity}
Noise fluctuations in the data can shift the inferred $\lt$ values, potentially biasing tidal and thus EoS inferences. This effect is particularly pronounced for detectors with lower sensitivity, where noise fluctuations can significantly influence the inferred NS parameters. In such cases, priors informed by EoSs can inadvertently impose constraints that are tighter than those supported by the data, effectively biasing the inference toward the properties of the chosen EoS ensemble.

To mitigate this issue, we introduce a controlled spread in the $\lt$ and $\dlt$ values derived from the EoS ensemble and scale it with detector sensitivity. This prevents the analysis from being artificially restricted by the EoS-informed prior when the data themselves are not sufficiently informative. As sensitivity improves with next-generation GW detectors, $\lt$ measurements will become more precise, and the likelihood will dominate the inference, allowing even relatively well-constrained EoS ensembles to be used safely.

A welcome consequence of this approach is that it also addresses a more fundamental limitation of EoS-informed priors. Even if the prior is broadened, one cannot guarantee that the true NS EoS lies within the ensemble used to construct it. The \emph{true} dense-matter EoS may lie inside the ensemble, at its boundaries, or entirely outside it. If the ensemble is too restrictive, the inference can become dominated by the prior, effectively “overfitting’’ to the chosen EoS set and introducing systematic biases. While not the primary motivation, incorporating a sensitivity-dependent spread also reduces the risk that the inference is biased by the particular EoS ensemble considered.

We use Fisher simulations to estimate the median uncertainty in $\lt$, denoted $\sigma_{\lt}$, as a function of the true $\lt$ of the signal. Since our focus is on the relatively low SNR regime, these simulations are performed using the current sensitivity LVK detector network, which includes the LIGO Hanford Observatory, LIGO Livingston Observatory, and the Virgo Observatory \cite{LIGOScientific:2014pky, VIRGO:2014yos}. Details of the simulation setup are provided in Appendix \ref{app1}. The resulting $\sigma_{\lt}(\lt)$ is interpolated over the same range of $\lt$ values used to construct the prior grid in \secname\ \ref{subsec:multid}.

For every $i^{\rm th}$ generated sample $(\mc^i, q^i, \lt^i, \dlt^i)$, we decide whether to perturb it by a single draw. \sk{The underlying EoS correlations should be preserved, so only a fraction of the samples are perturbed while the rest retain their exact EoS values. This is a standard approach in statistics, known as Huber contamination, in which only a fraction of a dataset is contaminated to build resilience against outliers \cite{Huber1964, Huber1965}. We contaminate a fraction $\varepsilon = 0.32$ of the samples, chosen to approximate the fraction expected to fall beyond $1\sigma$ under the measured uncertainty. A given sample is either perturbed or left at its exact EoS values; the same decision applies to both tidal parameters, so a sample is either noisy in both $\lt$ and $\dlt$ or exact in both.} For the perturbed samples, the spread in $\lt$ is set by $\sigma_{\lt}(\lt)$, and no spread is introduced in $\mc$, which is well constrained with uncertainties at the level of $\mathcal{O}(10^{-5})~\rm M_{\odot}$ \cite{LIGOScientific:2025slb}.

\sk{We introduce independent noise in $\dlt$ as well. There is no physical reason for detector noise to perturb $\lt$ and $\dlt$ in a correlated manner, so we do not tie the two fluctuations together. Applying the Fisher information to $\dlt$ yields uncertainties that are consistently large and nearly independent of the component masses or $\lt$, contrary to the case for $\sigma_{\lt}(\lt)$. We therefore assign $\dlt$ a fixed Gaussian spread with a standard deviation of 500, rather than a binary dependent Fisher error. This reflects the weak measurability of $\dlt$ at current detector sensitivity while keeping the prior broad enough not to over-constrain it.}

This sensitivity-motivated broadening accounts for the finite resolving power of the GW detector and helps reduce biases or overfitting to a specific EoS ensemble. The spread shown in \figname~\ref{fig:3_prior_comparison} arises directly from this procedure.

\sk{For the contaminated samples, large values of $\sigma_{\lt}$ can occasionally produce negative draws of $\lt$ values, which are unphysical. Rather than discarding these draws, we reflect them into the region $\lt \geq 0$, following the standard treatment in the literature. Reflection preserves the number of samples and redistributes them smoothly across the low $\lt$ region. No such treatment is required for $\dlt$, which has no strict requirements of being positive. }

\sk{The constraint $\Lambda_{1,2} \geq 0$ reshapes the prior, as shown in \secname~\ref{subsec:support} and \figname~\ref{fig:1_lt_support}. Samples drawn from the EoS without added noise are unaffected, since they are already physical with positive-definite $\Lambda_{1,2}$. The contaminated samples, however, can take unphysical values. To prevent the constraint from distorting the prior into an implied-prior shape, as happens for the uniform prior, we correct the contaminated samples analytically using the known form of the constraint. This preserves the intended shape of the prior throughout the analysis.}

\sk{One might ask why this construction is preferable to simply using the uniform prior or the noiseless EoS-based prior. The uniform prior lacks the EoS-based correlations, while the noiseless EoS prior can be overly restrictive. The contamination-and-correction treatment described here occupies the middle ground, and its benefit is evident in the results of \secname\ \ref{sec:results}, particularly for $\lt$ values near the edge of the EoS ensemble. }
\subsection{Multi-dimensional Parameter Correlations}
\label{subsec:multid}

Using the prior constructed by combining the methods described in \secname~\ref{subsec:eos} and \secname~\ref{subsec:sensitivity}, we build a fully EoS-informed four-dimensional conditional prior over the masses and tidal parameters, $(\mc,q,\lt,\dlt)$. This captures the physical correlations among all four parameters and significantly constrains the prior volume.

\begin{figure*}
    \centering
    \includegraphics[width=\linewidth]{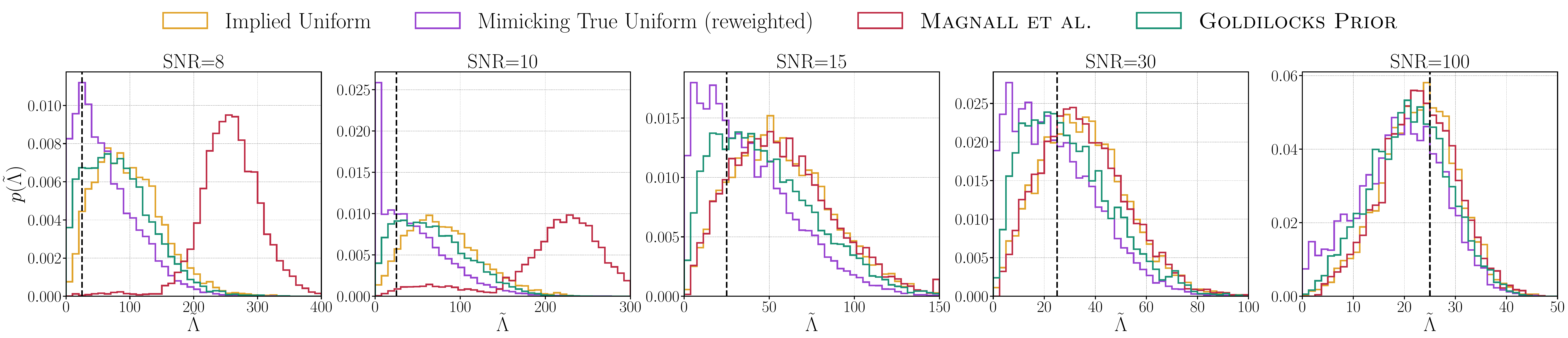}
    \caption{\sk{Posterior distributions for a symmetric $1.4$-$1.4\ \msol$ BNS with $\lt_{\rm inj} = 25$, injected across network SNRs of $8, 10, 15, 30, 100$ (left to right). The teal curve corresponds to the {\priorname} introduced in this work, orange shows the implied uniform prior of  \secname~\ref{subsec:support}, purple to the uniform posterior reweighted to mimic a true uniform prior described in \secname~\ref{subsec:reweighting} and the red curve to the priors prescribed by \citet{Magnall:2025zhm}. The dashed vertical line marks the injected value. For this configuration the lowest $\lt$ predicted by any EoS in the considered ensemble is $\sim67.4$, so the injected value lies well outside the ensemble. The \priorname\ contains the injected value at every SNR and its mode coincides with the injection from SNR 10 onward. On the contrary, the implied uniform and \citet{Magnall:2025zhm} priors are positively biased and the reweighted posterior shows spurious structure at low $\lt$ driven by the amplification of a small number of samples (see Appendix~\ref{app:reweighting})}}
    \label{fig:4_lt25_symm}
\end{figure*}

Under the constraint $m_1 \geq m_2$, one would generally expect $\Lambda_1 \leq \Lambda_2$. However, this ordering is not guaranteed in all scenarios. For example, NSs that might have undergone quantum chromodynamic phase transitions or contain exotic components, such as dark matter admixtures, may violate this relation (see Refs.\,\cite{Zhao:2018nyf}, \cite{Leung:2022wcf} and references therein). As the noisy $\lt$ and $\dlt$ values receive independent noise, the prior can occasionally admit $\Lambda_1 > \Lambda_2$, naturally avoiding assumptions about a specific ordering in $\Lambda_{1,2}$. A benefit of this framework is that it makes the prior flexible enough to accommodate even exotic compact object systems, should they be observed.

We construct a four-dimensional conditional prior in the $(\mc, q, \lt, \dlt)$ parameter space. The prior is evaluated on a fixed grid with dimensions $[128, 128, 128, 128]$. The $\mc$ grid spans values corresponding to BNSs with component masses between $1~\mathrm{M}_{\odot}$ and $2.5~\mathrm{M}_{\odot}$, covering the full range of plausible BNS configurations. The range of $\lt$ is set to $[0, 5000]$ to remain consistent with the uniform priors used for comparison. Similarly, $\dlt$ is evaluated over the range $[-1000, 1000]$. This grid is used to sample the prior shown in \figname~\ref{fig:3_prior_comparison}. For the full ensemble of $105{,}000$ EoSs, the grid evaluation requires only a few minutes and can be efficiently recomputed if needed.

For current-generation networks, the $\mc$ range used for analyses is generally narrower than the one used to create the grid.  This is because matched-filter searches provide a reasonably good estimate of $\mc$, particularly for BNS signals, due to the high density of templates in this region of parameter space \cite{Ewing:2023qqe}. As the grid evaluation is not computationally expensive, we create a denser grid in $\mc$ as and when required for our mock simulations. When we refer to $\mc$ here, we are specifically referring to the source frame chirp mass. For further details on how the source frame chirp mass is handled, readers are directed to Appendix~\ref{app:detector_source_frame}.

\section{Results}
\label{sec:results}
In this section, we present the results of several mock simulations designed to assess the performance of our proposed prior. We simulate GW signals, inject them into a hypothetical realisation of the current-sensitivity GW detector network, and perform parameter estimation to compare four priors: the implied uniform prior of \secname~\ref{subsec:support}, the uniform posterior reweighted to mimic a true uniform prior following \secname~\ref{subsec:reweighting}, the prior of \citet{Magnall:2025zhm}, and the \priorname\ introduced in this work. We use the files made public by \citet{Magnall:2025zhm} to compare with their prior as well \cite {Magnall:2025zhm,Altiparmak:2022bke, Ecker:2022dlg}.

\sk{A central result of this section is that the implied uniform prior cannot be used directly in the low-$\lt$ regime, where it is biased by construction. Its posteriors require correction, and reweighting is the standard means of applying one. As hinted at before in \secname~\ref{subsec:reweighting}, we show in this section decisively that reweighting cannot be relied upon in the low $\lt$ regime.}

\sk{Owing to the EoS ensemble used to construct the prior in \citet{Magnall:2025zhm}, it assigns negligible support to $\lt$ values that lie outside it. As a result, an event at such a value is rejected regardless of what the data prefer. Neither the implied uniform prior, nor its reweighted counterpart, nor the prior of \citet{Magnall:2025zhm} is consistent across the cases considered below as  each succeeds in some regimes and fails in others. The \priorname\ is consistent throughout, though not unbiased in every case. Where a residual bias remains, it traces to the measurability of the mass ratio and the tidal parameters rather than to the prior, as we demonstrate below.}

\sk{We organise the analyses by increasing dimensionality. The symmetric and asymmetric systems of \secname~\ref{subsec:symmm} and \secname~\ref{subsec:asymm} are analysed in the four-dimensional $(\mc, q, \lt, \dlt)$ space, with all remaining parameters fixed at their true values. This is the most favourable setting for the comparison priors. This is because with every other correlation removed, the prior-volume reduction offered by the \priorname\ cannot contribute, and the width of each posterior is set by the SNR and the detector sensitivity alone. The spinning systems of \secname~\ref{subsec:chi} extend this to six dimensions. A full treatment including extrinsic parameters is presented in the companion paper applying the \priorname\ to real events.}

All analyses are carried out using the Bayesian inference library \texttt{bilby} \cite{Romero-Shaw:2020owr, Ashton:2018jfp, lalsuite, swiglal} with the \texttt{dynesty} sampler \cite{Speagle:2019ivv} using $2048$ live points. To accelerate likelihood evaluations, we employ the relative binning implementation available within \texttt{bilby} \cite{Cornish:2010kf, Cornish:2021lje, Zackay:2018qdy}. Since our prior is evaluated only once and subsequently interpolated during sampling, this work represents the first application of such multi-dimensional joint priors in \texttt{bilby} analyses. This implementation yields computation speeds comparable to those of analytic priors. 

We also note that, since we aim to assess the performance of \priorname, all parameter estimation analyses in this work are performed as zero-noise injections. As shown in \citep{Vallisneri:2007ev, Rodriguez:2013oaa}, the posterior obtained from a zero-noise injection approximates the ensemble-averaged posterior over Gaussian noise realisations, allowing us to avoid the computational cost of repeated simulations while still capturing the expected measurement uncertainties in various parameters (also see  Refs.~\citep{Nissanke:2009kt, Wade:2014vqa}). One could carry out the same study over multiple noisy realisations to obtain the same ensemble behaviour; we instead use the zero-noise case as a computationally efficient representation of this average behaviour to benchmark the \priorname.

To demonstrate both the flexibility and robustness of the \priorname, we subject it to a series of stress tests in scenarios where parameter inference is expected to be particularly susceptible to bias. In \secname\ \ref{subsec:symmm}, we analyse a symmetric $1.4$--$1.4~\mathrm{M}_{\odot}$ BNS system, varying both the \sk{SNR} and the injected $\lt$ values. In \secname\ \ref{subsec:asymm}, we perform an analogous study for two asymmetric mass systems. In \secname\ \ref{subsec:chi}, we evaluate the performance of the priors for non-spinning and spinning BNS systems. Finally, in \secname\ \ref{subsec:chirpmass} and \secname\ \ref{subsec:deltalam}, we evaluate the accuracy of the inferred $\mc$ and $\dlt$ posteriors as well.
\subsection{Symmetric mass tests}
\label{subsec:symmm}

\sk{We begin with a symmetric $1.4$--$1.4~\rm M_{\odot}$ system. For this configuration the lowest $\lt$ predicted by any EoS in our ensemble is $67.4$, so an injection at $\lt_{\rm inj} = 25$ lies well outside the ensemble, and $\lt_{\rm inj} = 50$ sits just below its boundary. These values test the priors precisely where an EoS-informed construction is most likely to fail. \figname~\ref{fig:4_lt25_symm} shows the recovered $\lt$ posteriors for $\lt_{\rm inj} = 25$ at network SNRs of 8, 10, 15, 30 and 100.}

\sk{At SNR 8, the leftmost panel, the signal sits at the threshold below which detections are not considered. This is the least informative case available. The reweighted posterior appears to perform well, peaking close to the injection. However, this behaviour is an artefact because the peak is built from a handful of raw samples amplified by more than an order of magnitude (Appendix~\ref{app:reweighting}). The implied uniform posterior closely resembles its prior, retaining almost no support at $\lt = 0$. The \priorname\ is comparably uninformative at this SNR, but contains the injected value within its support. The prior of \citet{Magnall:2025zhm} places its mode near $\lt \approx 250$ and assigns negligible support at the injection, so an event of this kind would be recovered incorrectly with that prior. The \priorname\ outperforms the prior of \citet{Magnall:2025zhm} for two reasons. Its EoS ensemble is broader, so it does not overestimate $\lt$ as the more restrictive ensemble of \citet{Magnall:2025zhm} does, and the detector-sensitivity noise gives it support at low $\lt$ where the injection requires it.}

\sk{At SNR 10 the reweighting failure becomes explicit. The reweighted posterior develops a sharp spike at $\lt \approx 0$, produced by a single raw sample in a bin where the sampler found almost none. Reweighting is therefore not trustworthy without a number of live points large enough to populate the low-$\lt$ region, which the implied uniform prior makes prohibitively difficult by construction. The posterior distribution inferred by using \priorname\ now peaks at the injected value. The posterior remains broad, but the prior-driven bias identified in \secname~\ref{subsec:support} is already absent. Both the posteriors inferred from implied uniform and \citet{Magnall:2025zhm} priors peak away from the injected value, and remain positively biased at this SNR.}

\sk{At SNR 15 and 30 the reweighted posterior remains irregular and its mode drifts away from the injection, showing that it recovers neither the correct density nor the correct peak. The \priorname\ keeps its mode at the injected value while the posterior narrows. The implied uniform and \citet{Magnall:2025zhm} priors remain positively biased. We emphasise that this comparison is deliberately favourable to the uniform priors as all parameters other than $(\mc, q, \lt, \dlt)$ are fixed at their true values and thus, the significantly helpful prior-volume reduction that the \priorname\ provides cannot contribute. Even under these conditions, the three comparison priors are biased.}

\sk{At SNR 100 the likelihood dominates and all priors are expected to agree. They do, with the exception of the reweighted posterior, which continues to show noise at low $\lt$ where it amplifies sparse samples. This persists even at the highest SNR considered, and is one of our arguments against using reweighted posteriors obtained from the implied uniform priors for BNS inference.} 

\sk{All four posteriors nonetheless retain a slight bias toward lower $\lt$ at this SNR. This is not a prior effect. For a symmetric injection the mass ratio rails against $q = 1$, so its posterior can only extend toward lower $q$, producing a one-sided tail. Since $\lt$ depends on the component masses through Eq.~\ref{eq:lt}, this tail maps directly onto lower $\lt$ values and draws the recovered posterior slightly below the injection. The effect follows from the definition of $\lt$ and the mass ratio boundary, and is common to all four priors.}

This result encapsulates the central motivation of the \priorname. EoS-informed correlations are a powerful tool for improving tidal parameter recovery, but only when the prior remains broad enough to reflect genuine physical uncertainty rather than suppress regions of parameter space that the data cannot rule out.

As the performance of a prior should be judged by its recovery across a range of possible $\lt$ values, we now vary the injected value at a fixed network SNR of $10$. \figname~\ref{fig:4_lt25_symm} fixed $\lt_{\rm inj}$ and varied the SNR. \figname~\ref{fig:5_all_lam} does the reverse, showing $\lt_{\rm inj} = 25$, $50$ and $100$. We use SNR $10$ because it represents the regime in which the priors still dominate the inference.

\begin{figure*}
    \centering
    \includegraphics[width=\linewidth]{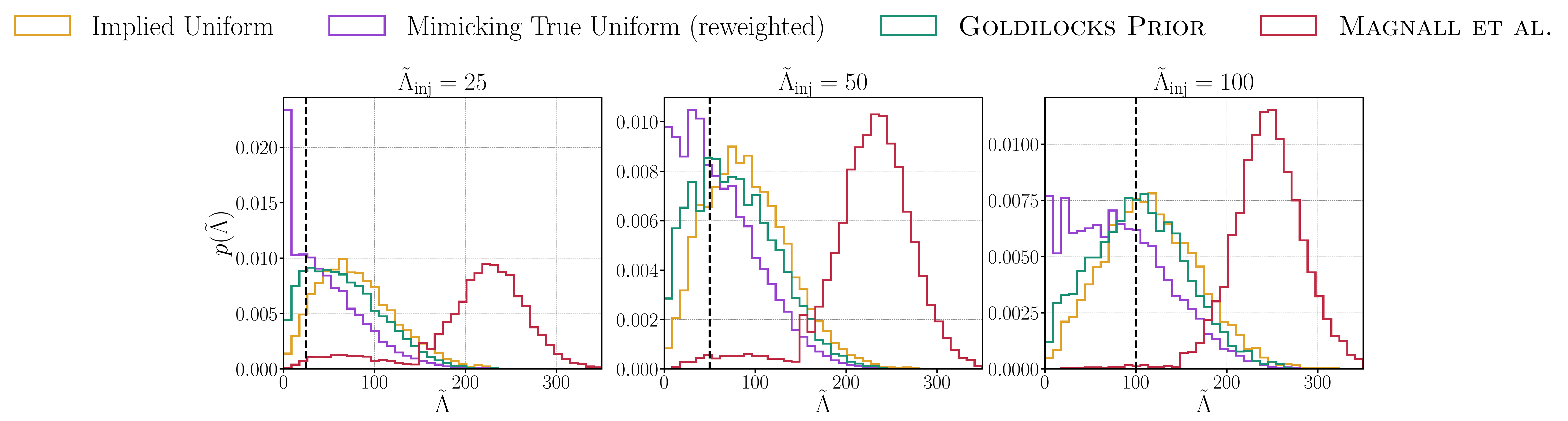}
    \caption{\sk{Posterior distributions of $\lt$ for a symmetric $1.4$--$1.4\ \msol$ BNS at a network SNR of $10$, for injected values $\lt_{\rm inj} = 25$, $50$ and $100$ (left to right). Curves follow \figname~\ref{fig:4_lt25_symm} with teal for the \priorname, orange for the implied uniform prior, purple for the reweighted uniform posterior, and red for the prior of \citet{Magnall:2025zhm}. The dashed vertical line marks the injected value. The lowest $\lt$ predicted by the EoS ensemble for this configuration is $67.4$, so the first two injections lie outside it and the third lies within reach. The \priorname\ places its mode at the injection in all three cases. The prior of \citet{Magnall:2025zhm} peaks near $\lt \approx 240$ irrespective of the injected value. The reweighted posterior shows spurious low-$\lt$ structure in all three panels, including at $\lt_{\rm inj} = 100$, where the  implied uniform posterior is already unbiased.}}
    \label{fig:5_all_lam}
\end{figure*}

\sk{The posterior recovered under the \priorname\ peaks at the injected value in all three panels. Under the implied uniform prior the posterior is positively biased at $\lt_{\rm inj} = 25$ and $50$, and the bias diminishes as the injected value increases, with $\lt_{\rm inj} = 100$ recovered accurately. This follows from the geometry described in \secname~\ref{subsec:support} rather than from any EoS information. The implied uniform prior suppresses low $\lt$ because few combinations of $\Lambda_{1,2}$ produce small $\lt$, and this suppression matters less as the injection moves away from that region.}

\sk{The inference using \citet{Magnall:2025zhm} is biased for a different reason. The posterior recovered under it peaks near $\lt \approx 240$ in all three panels, moving by only a few tens of units across a factor of four in $\lt_{\rm inj}$. Its EoS ensemble does not support any of the injected values, so the recovered posterior reports the ensemble rather than the data. This bias would weaken for injections inside the region its ensemble supports for a $1.4$--$1.4~\rm M_{\odot}$ system, which none of the three values considered here reach.}

\sk{The reweighted posterior is irregular in all three panels and introduces support at low $\lt$ in each. The $\lt_{\rm inj} = 100$ panel is the clearest case. There the implied uniform posterior is already unbiased and no correction is needed, yet reweighting assigns substantial weight to a low-$\lt$ region the data do not favour. The correction does not merely fail to improve the posterior, it degrades one that was already correct.}

\begin{figure*}[t]
    \centering
    \refstepcounter{figure} % Forces the counter to Figure 6 NOW
    
    \begin{minipage}{\textwidth}
        \centering
        \includegraphics[width=\linewidth]{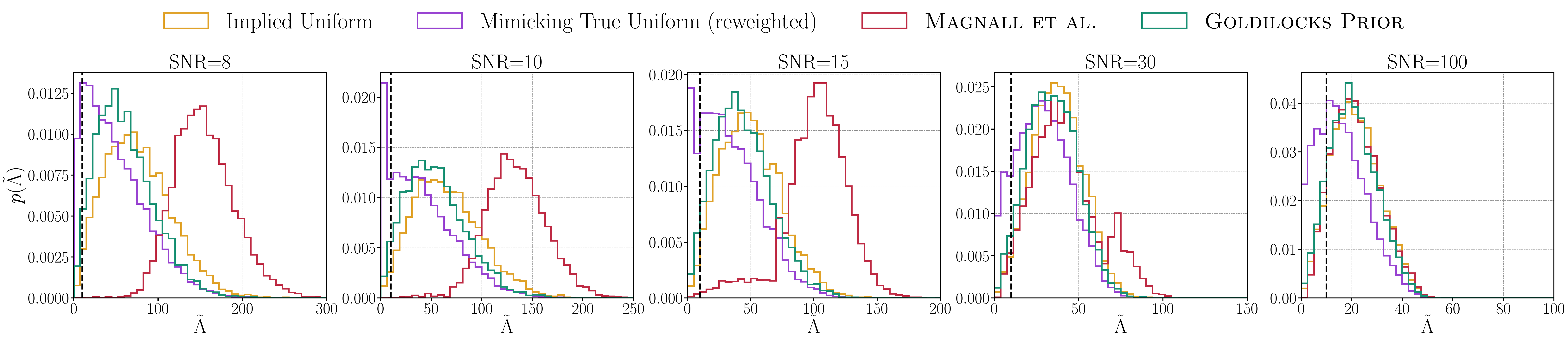}
        \\ \small (a)
        \makeatletter \def\@currentlabel{\thefigure(a)} \makeatother \label{fig:6_masym}
    \end{minipage}

    \vspace{0.5cm}

    \begin{minipage}{\textwidth}
        \centering
        \includegraphics[width=\linewidth]{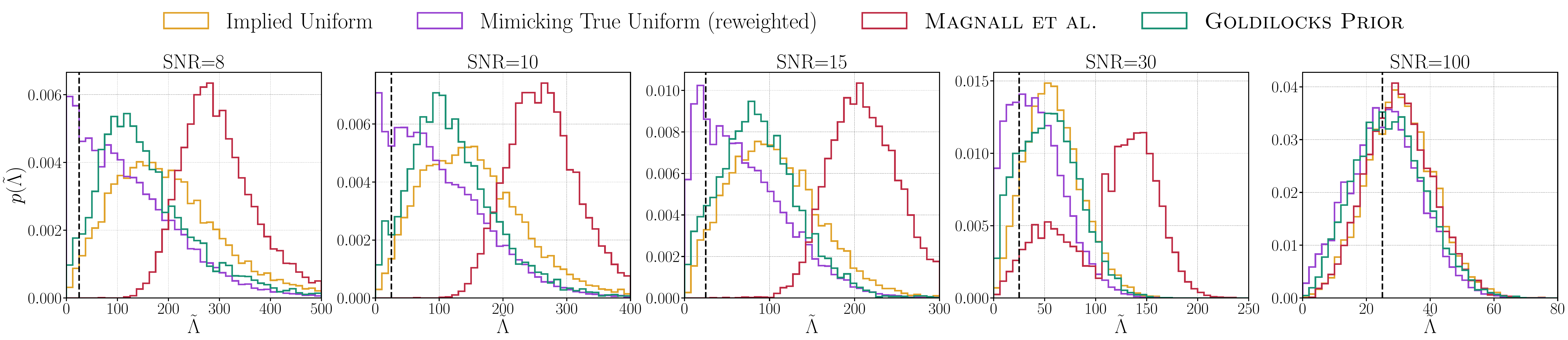}
        \\ \small (b)
        \makeatletter \def\@currentlabel{\thefigure(b)} \makeatother \label{fig:6_hasym}
    \end{minipage}
    
    \addtocounter{figure}{-1} 
    \caption{\sk{Recovered $\lt$ posteriors for the two asymmetric systems at network SNRs of 8, 10, 15, 30 and 100 (left to right). Panel (a) shows the intermediate asymmetry case, $1.8$--$1.6~\rm M_{\odot}$ with $\lt_{\rm inj} = 10$, for which the lowest $\lt$ predicted by the EoS ensemble is $17.7$. Panel (b) shows the high asymmetry case, $1.2$--$2.0~\rm M_{\odot}$ with $\lt_{\rm inj} = 25$. Curves follow \figname~\ref{fig:4_lt25_symm}. The dashed vertical line marks the injected value. In (a) the recovered posteriors remain biased at SNR 100, since the mass ratio cannot be resolved away from the boundary at $q = 1$. In (b) the mass ratio is far enough from that boundary for the data to recover it, and the posteriors converge toward the injection. The corresponding medians and $90\%$ credible intervals are listed in \tabname~\ref{tab:asym}.}}
    \label{fig:6_asym}
\end{figure*}

\sk{This is the behaviour anticipated in \secname~\ref{subsec:reweighting}. The weights are largest exactly where the implied uniform prior sampled least, so the reweighted density in that region is set by a handful of points rather than by the likelihood. Across the three injections, neither the uncorrected uniform posterior nor its reweighted counterpart is reliable. The uncorrected posterior is biased at low $\lt$ and unbiased at $\lt_{\rm inj} = 100$, while the reweighted posterior is irregular throughout. Only the \priorname\ recovers the injected value consistently.}

\subsection{Asymmetric mass tests}
\label{subsec:asymm}
We can also consider signals like GW190425, which involve asymmetric masses \cite{LIGOScientific:2020aai}. As discussed in \citet{Khadkikar:2025awf}, hypotheses that GW190425 originates from an NSBH merger or a binary black hole merger must be considered seriously, in addition to the widely considered BNS scenario. With a prior like the \priorname, which correctly treats the low-$\lt$ region, it is particularly interesting to analyse such systems, as they can exhibit very low $\lt$ values due to the presence of one or two black holes in the binary. To test the performance of our prior set for asymmetric mass systems, we consider two cases:
\begin{enumerate}
    \item \sk{Intermediate asymmetry: $1.8$--$1.6~\rm M_{\odot}$ system}
    \item \sk{High asymmetry: $1.2$--$2.0~\rm M_{\odot}$ system}
\end{enumerate}
\sk{A detailed analysis of GW190425 itself is presented in the companion paper applying the \priorname\ to real events.}

\begin{table*}[t]
\centering
\caption{\sk{Recovered $q$ and $\lt$ for the two asymmetric systems, quoted as median with $90\%$ credible interval. The intermediate case has $q_{\rm true} = 0.889$ and $\lt_{\rm inj} = 10$, the high asymmetry case $q_{\rm true} = 0.6$ and $\lt_{\rm inj} = 25$.}}
\label{tab:asym}
\begin{ruledtabular}
\begin{tabular}{llcccc}
 &  & \multicolumn{2}{c}{Intermediate asymmetry, $1.8$--$1.6~\rm M_{\odot}$} & \multicolumn{2}{c}{High asymmetry, $1.2$--$2.0~\rm M_{\odot}$} \\
\cline{3-4}\cline{5-6}
SNR & Prior & $q$ & $\lt$ & $q$ & $\lt$ \\
\hline
8   & Implied Uniform         & 0.929 [0.828, 0.993] & 89.4 [22.5, 1843.6] & 0.647 [0.600, 0.879] & 198.0 [51.4, 1970.2] \\
    & \citet{Magnall:2025zhm} & 0.952 [0.874, 0.995] & 151.9 [101.4, 218.0] & 0.676 [0.633, 0.747] & 289.5 [195.8, 444.5] \\
    & \priorname              & 0.932 [0.835, 0.993] & 57.0 [14.4, 125.4] & 0.632 [0.595, 0.708] & 137.9 [33.4, 352.8] \\
\hline
10  & Implied Uniform         & 0.937 [0.844, 0.994] & 65.5 [18.6, 1178.7] & 0.631 [0.598, 0.694] & 148.1 [42.1, 337.8] \\
    & \citet{Magnall:2025zhm} & 0.957 [0.886, 0.996] & 132.7 [89.1, 184.4] & 0.664 [0.631, 0.711] & 261.8 [176.9, 373.6] \\
    & \priorname              & 0.937 [0.845, 0.993] & 51.7 [12.7, 109.9] & 0.623 [0.595, 0.669] & 112.6 [26.1, 252.3] \\
\hline
15  & Implied Uniform         & 0.944 [0.866, 0.994] & 48.3 [14.8, 92.7] & 0.618 [0.597, 0.647] & 100.8 [27.4, 197.4] \\
    & \citet{Magnall:2025zhm} & 0.965 [0.901, 0.997] & 103.6 [49.3, 137.8] & 0.647 [0.625, 0.674] & 209.8 [149.5, 287.5] \\
    & \priorname              & 0.942 [0.862, 0.994] & 41.7 [11.4, 85.1] & 0.614 [0.595, 0.640] & 84.9 [18.6, 170.7] \\
\hline
30  & Implied Uniform         & 0.945 [0.882, 0.995] & 37.2 [13.5, 62.3] & 0.607 [0.597, 0.620] & 57.2 [18.1, 108.1] \\
    & \citet{Magnall:2025zhm} & 0.951 [0.884, 0.995] & 39.8 [13.8, 79.4] & 0.623 [0.600, 0.637] & 125.2 [29.5, 174.4] \\
    & \priorname              & 0.944 [0.881, 0.995] & 34.2 [11.1, 59.9] & 0.606 [0.595, 0.620] & 53.7 [11.2, 104.5] \\
\hline
100 & Implied Uniform         & 0.908 [0.882, 0.959] & 21.1 [6.9, 38.4] & 0.601 [0.597, 0.605] & 30.4 [13.5, 48.4] \\
    & \citet{Magnall:2025zhm} & 0.908 [0.883, 0.959] & 20.9 [7.7, 38.5] & 0.601 [0.597, 0.605] & 30.0 [14.0, 47.9] \\
    & \priorname              & 0.906 [0.881, 0.955] & 20.0 [6.4, 36.9] & 0.600 [0.596, 0.605] & 27.1 [10.0, 46.8] \\
\end{tabular}
\end{ruledtabular}
\end{table*}
\begin{figure*}[t]
    \centering
    \refstepcounter{figure} % Forces the counter to 7 NOW

    \begin{minipage}{\textwidth}
        \centering
        \includegraphics[width=\linewidth]{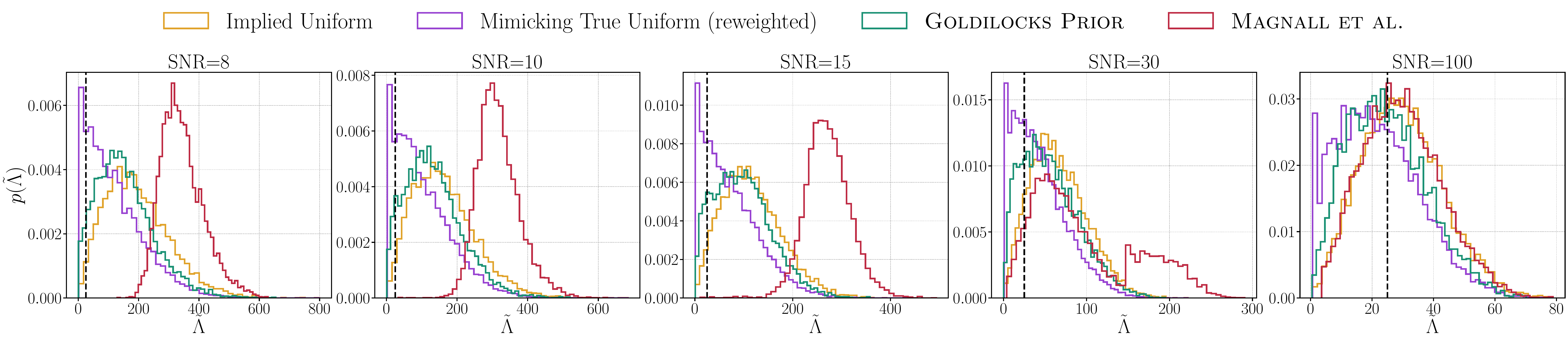}
        \\ \small (a)
        \makeatletter \def\@currentlabel{\thefigure(a)} \makeatother \label{fig:7_lowspin}
    \end{minipage}

    \vspace{0.5cm}

    \begin{minipage}{\textwidth}
        \centering
        \includegraphics[width=\linewidth]{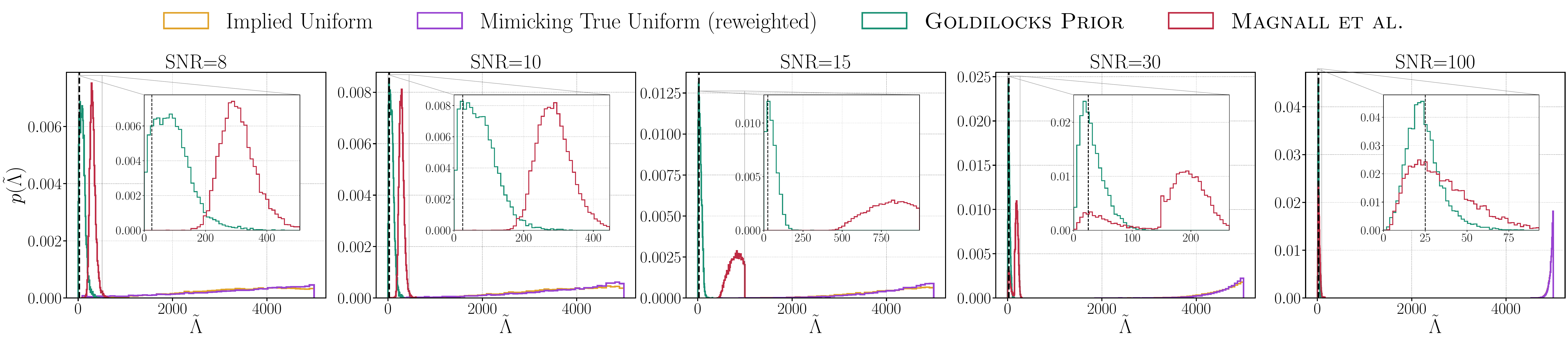}
        \\ \small (b)
        \makeatletter \def\@currentlabel{\thefigure(b)} \makeatother \label{fig:7_highspin}
    \end{minipage}
    
    \addtocounter{figure}{-1} % Subtracts 1 so the \caption command doesn't make it Fig 8
\caption{\sk{Recovered $\lt$ posteriors for the symmetric $1.4$--$1.4~\rm M_{\odot}$ system with spins sampled, at network SNRs of 8, 10, 15, 30 and 100 (left to right), both at $\lt_{\rm inj} = 25$. Panel (a) shows the low-spin case, $\chi_i \leq 0.1$. Panel (b) shows the high-spin case, $\chi_i = 0.5$. Curves follow \figname~\ref{fig:4_lt25_symm}. The dashed vertical line marks the injected value. In the low-spin case, the recovered posteriors match the non-spinning results of \secname~\ref{subsec:symmm}. In the high-spin case, the \priorname\ recovers the injection across the full SNR range, while the implied uniform and reweighted posteriors rail to the upper edge of the prior at $\lt = 5000$ and the prior of \citet{Magnall:2025zhm} sits in a separate high mode. The full set of parameters for the high-spin case is shown in \figname~\ref{fig:appendix_highspin}.}}
    \label{fig:7_spins}
\end{figure*}
\sk{We begin with the intermediate case. The $1.8$--$1.6~\rm M_{\odot}$ system closely resembles GW190425, which motivates its choice here. The lowest $\lt$ predicted by the EoS ensemble for this configuration is $17.7$, and we inject $\lt_{\rm inj} = 10$, which lies below that bound. \figname~\ref{fig:6_masym} shows the recovered posteriors at network SNRs of 8, 10, 15, 30 and 100, and \tabname~\ref{tab:asym} lists the corresponding medians and $90\%$ credible intervals.}

\sk{At SNR 8, the \priorname\ recovers $\lt = 57.0\,[14.4, 125.4]$, while the implied uniform prior gives $89.4\,[22.5, 1843.6]$ and \citet{Magnall:2025zhm} gives $151.9\,[101.4, 218.0]$. The implied uniform interval is very broad, reflecting a second mode at high $\lt$ that lies outside the range plotted in \figname~\ref{fig:6_masym}. The posteriors at SNR 10 are qualitatively similar. Across both SNRs the reweighted posterior peaks nearest the injected value and appears to recover it best of the four. This agreement is not a consequence of the data, and understanding the reason behind it requires analysing the inferred $q$ posterior.}

\sk{The true mass ratio for this system is $q = 1.6/1.8 \approx 0.889$. As \tabname~\ref{tab:asym} shows, the recovered $q$ is biased toward unity at every SNR, with the injected value lying inside the $90\%$ interval but near its lower edge in several cases. Sampling uniformly in $m_i$ favours symmetric configurations, biasing $q$ toward unity, and because the true value already lies close to the boundary, the posterior additionally rails against the hard limit at $q = 1$. Both effects make the system appear more symmetric than it is unless the SNR is high enough to resolve the posterior away from the boundary. Through the dependence of $\lt$ on $m_i$ in Eq.~\ref{eq:lt}, this bias toward higher $q$ pushes $\lt$ above the injected value. The effect persists to SNR 100, where the implied uniform, \citet{Magnall:2025zhm}, and \priorname\ posteriors converge on $\lt \approx 21$, roughly twice the injection. It is common to all three priors, and unbiased recovery of $\lt$ for slightly asymmetric systems requires progressively higher SNR.}

\sk{The reweighted posterior is the exception, and the reason exposes a limitation of the method. The $q$ prior is identical across all three full Bayesian analyses and has support throughout the relevant range, so all three recover qualitatively similar $q$ posteriors from the data, as shown in \tabname~\ref{tab:asym}. Reweighting is a post-processing step applied afterwards, and its importance weight is a function of both $\lt$ and $q$. As it lowers $\lt$, it lowers $q$ in tandem, pulling the mass ratio below the value the three analyses agree on and opposite to the prior-driven bias toward $q = 1$ present in all of them. We conclude that $q$ had already been faithfully sampled, and reweighting displaces it.}

\sk{Reweighting can therefore move both $\lt$ and $q$ in ways not supported by the data. Its downward shift in $\lt$ happens to counter the upward bias present in the other three priors, so the reweighted $\lt$ posterior peaks near the injected value. This apparent accuracy is a byproduct of displacing the $q$ posterior that had already been faithfully sampled, not a faithful recovery by itself. The method is therefore unreliable in this regime, independent of whether it happens to agree with the injected $\lt$ in a given case.}

\sk{The high asymmetry case shows the same ranking of the priors, but with a different high-SNR outcome. For the $1.2$--$2.0~\rm M_{\odot}$ system, the true mass ratio is $q = 0.6$, far from the boundary at unity, and we inject $\lt_{\rm inj} = 25$. \figname~\ref{fig:6_hasym} shows the recovered posteriors. At low SNRs, the $q$ posterior is again biased toward higher values, because sampling uniformly in $m_i$ favours symmetric configurations and pushes $q$ toward unity regardless of the true value. Through Eq.~\ref{eq:lt} this displaces $\lt$ upward, and the priors rank as before, with the reweighted posterior peaking nearest the injection and the other three biased high. At SNR 8, the \priorname\ recovers $\lt = 137.9\,[33.4, 352.8]$, against $198.0\,[51.4, 1970.2]$ under the implied uniform prior and $289.5\,[195.8, 444.5]$ under \citet{Magnall:2025zhm}, and is the closest of the four to the injection.}

\sk{Unlike the intermediate case, this system resolves as the SNR increases. The recovered $q$ falls from $0.632$ at SNR 8 to $0.600\,[0.596, 0.605]$ at SNR 100, an interval that contains the true value, and $\lt$ follows, reaching $27.1\,[10.0, 46.8]$ under the \priorname\ against the injection of $25$. As the true mass ratio lies well away from the boundary, the prior preference for symmetry is not reinforced by a rail against $q = 1$, and the data recover $q$ once the signal is sufficiently loud. The reweighted posterior again peaks near the injection at low SNR, but this reflects the same downward displacement of $q$ and $\lt$ under reweighting, not a faithful recovery.}
\subsection{Spinning system tests}
\label{subsec:chi}

\begin{figure*}
    \centering
    \includegraphics[width=\linewidth]{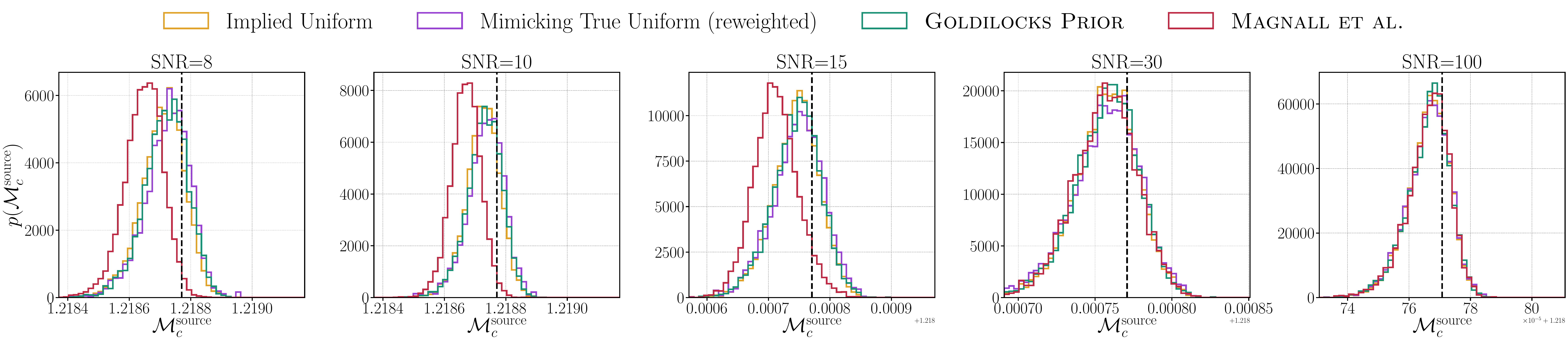}
    \caption{\sk{Source-frame $\mc$ posteriors for a symmetric $1.4$--$1.4~\rm M_{\odot}$ BNS injection with $\lt_{\rm inj} = 100$, at network SNRs of 8, 10, 15, 30 and 100 (left to right). Curves follow \figname~\ref{fig:4_lt25_symm}. The dashed vertical line marks the injected value. At SNR 30 and 100 all four priors recover $\mc$ without significant bias. At the three lower SNRs the prior of \citet{Magnall:2025zhm} is shifted toward lower $\mc$, following from the $\lt$ overestimation seen in \figname~\ref{fig:5_all_lam} and the correlation between the two parameters. This highlights the importance of using physically informed priors that maintain unbiased inference in both $\lt$ and correlated parameters like $\mc$.}}
    \label{fig:8_mc}
\end{figure*}

\begin{figure*}
    \centering
    \includegraphics[width=\linewidth]{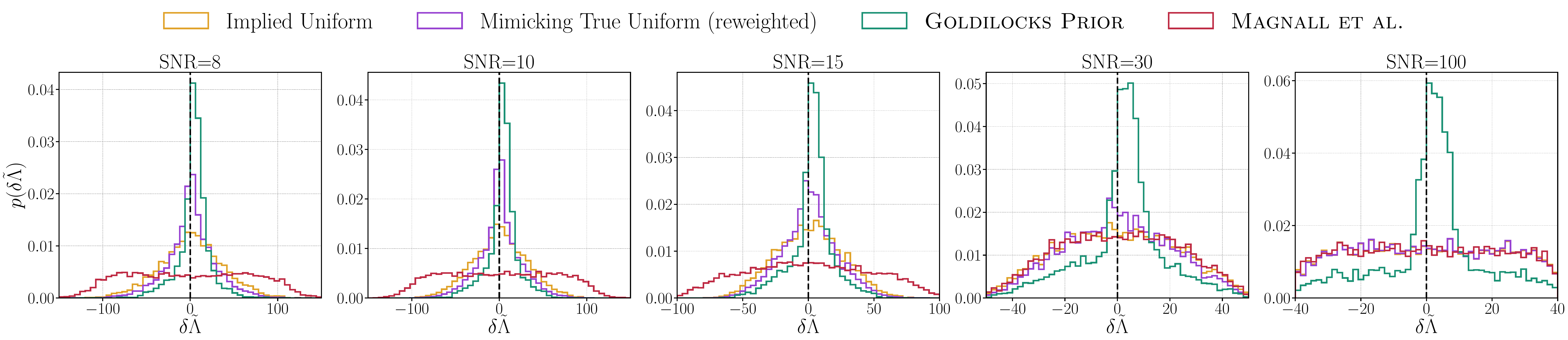}
    \caption{\sk{Posterior distributions of $\dlt$ for the symmetric $1.4$--$1.4~\rm M_{\odot}$ BNS injection with $\lt_{\rm inj} = 100$, at network SNRs of 8, 10, 15, 30 and 100 (left to right). Curves follow \figname~\ref{fig:4_lt25_symm}. The dashed vertical line marks the injected value, $\dlt = 0$. The \priorname, which treats $\dlt$ as a dimension of the joint prior, produces the narrowest posterior at every SNR. The implied uniform prior, its reweighted counterpart, and the prior of \citet{Magnall:2025zhm} all recover a posterior peaked at zero, but substantially broader.}}
    \label{fig:9_dlt}
\end{figure*}
\sk{The analyses so far have held the component spins fixed. Spin enters the waveform at lower post-Newtonian order than the tidal terms, and can influence the recovery of $\lt$ through the spin-induced quadrupole moment \cite{Poisson:1995ef, Poisson:1997ha}. We therefore repeat the symmetric $1.4$--$1.4~\rm M_{\odot}$ study with the spins sampled, extending the parameter space to the six dimensions $(\mc, q, \lt, \dlt, \chi_1, \chi_2)$. We consider two cases. The first uses a low-spin prior with aligned component spins restricted to  $\chi_i \leq 0.1$, twice the width of the standard low-spin prior of \secname~\ref{sec:LVK_priors}, so that the expected spins of merging BNSs lie well inside the sampled range; the injected spins are $\chi_i = 0.01$. The second injects $\chi_i = 0.5$ and samples under a high-spin prior allowing $\chi_i \leq 0.89$.}

\sk{\figname~\ref{fig:7_lowspin} shows the low-spin case at $\lt_{\rm inj} = 25$, across the same network SNRs as \figname~\ref{fig:4_lt25_symm}, so the effect of adding spin can be compared directly against the four-dimensional inference presented earlier. The recovered posteriors are qualitatively indistinguishable from the non-spinning results of \secname~\ref{subsec:symmm}. The low-spin prior confines the component spins to a narrow range, so the two additional dimensions open few new configurations and leave the tidal recovery essentially unchanged. The \priorname\ remains closest to the injection across the sweep, while the reweighted posterior again develops spurious support at low $\lt$ and remains rough even at SNR 100. The behaviour established for the non-spinning symmetric system therefore carries over directly once low spins are included.}

\sk{Panel (b) shows the high-spin case, with both components at $\chi_i = 0.5$. Here the differences between the priors are at their most pronounced in all considered cases. The implied uniform and reweighted posteriors fail to recover $\lt$ at any SNR, railing instead to the upper edge of the prior near $\lt = 5000$, while the \priorname\ holds the injection throughout and the prior of \citet{Magnall:2025zhm} sits in a separate high mode. The failure of the uniform priors is not confined to $\lt$. As shown in \figname~\ref{fig:appendix_highspin}, every parameter is recovered incorrectly together.}

\sk{$\mc$ is the leading-order parameter, and when it is incorrectly recovered, the rest of the parameters are all affected due to correlations. For the uniform priors, the recovered $\mc$ is biased high, corresponding to a heavier and therefore shorter signal. As $\mc$ and $\chi_i$ are inversely correlated, compensating for the too-heavy $\mc$, the sampler drives the spins strongly negative as shown in \figname~\ref{fig:appendix_highspin}. Similar correlations drive  $\lt$ to large values, and with no EoS information to bound the tidal deformability, $\lt$ runs to an arbitrary high value at the prior boundary. $q$ is correspondingly pulled toward high asymmetry. The result is a self-consistent but entirely unphysical solution across all six parameters.}

\sk{The EoS-informed priors avoid this. Both the \priorname\ and the prior of \citet{Magnall:2025zhm} couple $\mc$ and $\lt$, so the combination of a heavy $\mc$ with an extreme $\lt$ is disfavoured, which anchors $\mc$ and keeps the remaining parameters in place. This is a clear case in which an EoS-informed prior is not merely an improvement but a requirement for the inference to converge on a physical solution. The \priorname\ again outperforms the prior of \citet{Magnall:2025zhm} in $\lt$, owing to its noise-broadened support at low $\lt$ and its broader EoS ensemble.}

\sk{We note that spins of $\chi_i = 0.5$ are far higher than expected for merging BNSs. This configuration is included as a stress test of the inference pipeline rather than as a realistic source. We strongly emphasise through these high-spin results that the pipeline itself fails in this regime unless the prior carries EoS information, independent of whether such a system is astrophysically likely.}

\subsection{Improvements in the chirp mass constraints}
\label{subsec:chirpmass}

We conclude from the results in \secname~\ref{subsec:symmm} and \secname~\ref{subsec:asymm} that the \priorname\ is the most robust prior for unbiased inference of $\lt$ across multiple scenarios. As $\mc$ and $\lt$ are inherently correlated, achieving unbiased $\lt$ constraints also reduces uncertainty in the $\mc$ posterior. To illustrate this effect, we present the source-frame $\mc$ posteriors for all priors, assuming $\lt_{\rm inj} = 100$, in \figname~\ref{fig:8_mc}. \sk{All chirp mass results in this work are quoted in the source frame; the conversion from the detector frame is described in Appendix~\ref{app:detector_source_frame}.}

For BNS GW signals, $\mc$ is the best-constrained parameter, and thus the posteriors on it are already narrow, as seen in \figname~\ref{fig:8_mc}. \sk{At SNR 30 and 100 all four priors recover $\mc$ without significant bias, consistent with the unbiased $\lt$ values recovered at those SNRs. At SNR 8, 10 and 15, however, the prior of \citet{Magnall:2025zhm} produces a considerably biased $\lt$ posterior, as seen in \figname~\ref{fig:5_all_lam}. This bias propagates to $\mc$, leading to an underestimated $\mc$ constraint corresponding to the overestimated $\lt$ value.} This demonstrates why broader, yet physically informed priors are important, not only for unbiased $\lt$ posteriors but also for their indirect effect on $\mc$ inference.
\subsection{Improvements in the reduced tidal deformability constraints}
\label{subsec:deltalam}

As highlighted in \secname~\ref{sec:LVK_priors}, $\dlt$ can be an important parameter for inference, as better measurement of $\dlt$ also improves the resolution of the individual $\Lambda_{1,2}$. \sk{\figname~\ref{fig:9_dlt} shows the $\dlt$ posteriors for the symmetric $1.4$--$1.4~\rm M_{\odot}$ injection with $\lt_{\rm inj} = 100$, across the SNR range considered.} Since \priorname\ is a \sk{four-dimensional} prior with a physically consistent treatment of $\dlt$, its posterior is the narrowest among all priors considered \sk{at every SNR}. The uniform and \citet{Magnall:2025zhm} priors also correctly recover the $\dlt$ value peaking at 0, but are significantly broader.
\section{Conclusions}
\label{sec:conclusions}
BNS GW detections are critically important not only for GW astrophysics but also for cosmology, nuclear physics, quantum chromodynamics, and many other related fields. As GWs constrain a unique parameter, the tidal deformability $\lt$, the constraints obtained from BNS observations are of paramount importance for both multimessenger analyses and theoretical studies based on these observations. Regardless of the type of BNS observed, systematic biases arising from the choice of priors can significantly impact downstream analyses that rely on GW constraints.

Uniform priors are widely used in the literature for standard BNS analyses. However, because these priors treat masses and tidal parameters independently, despite their inherent correlations, and largely disregard the $\dlt$ parameter, they can lead to unphysical results and biased estimates of NS properties. Similarly, other studies have proposed physics-informed priors based on EoS ensembles. Due to the improper treatment of $\dlt$ and the strong influence of the chosen EoS ensemble, these approaches can also produce biased results. \sk{We demonstrate these limitations, along with the unreliability of correcting uniform-prior posteriors by reweighting, in detail in \secname~\ref{sec:LVK_priors}.}

Our approach toward an unbiased prior for BNS analyses focuses on incorporating physical information from EoSs while ensuring that the data itself remains informative. We account for detector sensitivity, the presence of noise, and the joint nature of tidal parameters, avoiding overconstraining the data with prior assumptions. In this work, we present \priorname, which integrates all of these considerations as described in \secname\ \ref{sec:newprior}.

\sk{To test the robustness of priors, we perform extensive mock BNS injections designed to challenge each prior, revealing where unphysical results or overly broad constraints may arise. We compare four priors, the implied uniform prior, its reweighted counterpart, the prior of \citet{Magnall:2025zhm}, and the \priorname, for a symmetric $1.4$--$1.4~\rm M_{\odot}$ system (\secname~\ref{subsec:symmm}), two asymmetric systems of $1.8$--$1.6$ and $1.2$--$2.0~\rm M_{\odot}$ (\secname~\ref{subsec:asymm}), and spinning systems (\secname~\ref{subsec:chi}), and we evaluate the effect on the $\mc$ and $\dlt$ posteriors (\secname~\ref{subsec:chirpmass} and \secname~\ref{subsec:deltalam}). Across these tests the \priorname\ is the only prior to recover $\lt$ without the boundary-driven bias of the implied uniform prior, the ensemble-driven bias of \citet{Magnall:2025zhm}, or the sampling artefacts of reweighting. The high-spin configuration is the most demanding case, where the uniform priors fail across every parameter and only the EoS-informed priors return a physical solution. The \priorname\ performs the best even in this extreme case.}

\sk{Some limitations persist independently of the chosen prior. For slightly asymmetric systems, the mass ratio rails against the boundary at $q = 1$, biasing $\lt$ towards higher values until the SNR is large enough to resolve it, an effect common to every prior used. Reweighting the uniform prior posterior is separately unreliable, since it can displace both $\lt$ and $q$ in ways the data do not support. Neither effect is remedied by the \priorname, but it alone avoids the prior-driven biases while retaining support in the low-$\lt$ region that the data may favour. By providing this for both $\lt$ and correlated parameters such as $\mc$ and $\dlt$, the \priorname\ establishes a robust framework for tidal parameter estimation in BNS analyses. Beyond this study using mock injections, we release a companion paper applying the \priorname\ to real BNS events, illustrating its utility in extracting reliable neutron star properties from observational data.}

This study has broad implications, including extensions to NSBH systems, the effect of future LVK upgrades and next-generation detectors, impacts on EoS model selection, and many others. We hope that the use of \priorname\ will be an important step toward unravelling the properties of the densest matter in the universe.

\begin{acknowledgments}
For insightful discussions and feedback, we thank K. G. Arun, Nils Andersson,  Jolien Creighton, Nathan Johnson-McDaniel and Jocelyn Read.  We also thank Colm Talbot for help in the development of the first joint prior templates implemented in \textsc{bilby}. This research is supported by National Science Foundation grants AST-2307147, PHY-2308886, PHY-2309064, PHY 2608011 and AST-2606673. 
\end{acknowledgments}
\bibliography{bilbiography}% Produces the bibliography via BibTeX.

\appendix
\section{Fisher Analysis}
\label{app1}

For this study, we focus on calculating the expected uncertainty in $\tilde{\Lambda}$, denoted $\sigma_{\tilde{\Lambda}}$, as a function of the true $\tilde{\Lambda}$ of the BNS system under analysis. Since this is inherently dependent on GW detector sensitivity, we estimate $\sigma_{\tilde{\Lambda}}$ in a systematic manner using the Fisher information formalism. Here, the aLIGO power spectral density curve constructed for O4 (\cite{LIGO_T980044_10}) is used to simulate the sensitivities of LIGO Hanford and LIGO Livingston, while the Advanced Virgo design sensitivity curve is used to simulate Virgo (\cite{LIGO_P1200087_v42}). For a detailed discussion of the Fisher approach in GW data analysis, the reader is referred to Refs.~\cite{Cutler:1994ys, Poisson:1995ef}.

It is not suitable to use an astrophysically motivated population to compute $\sigma_{\lt}$, as realistic astrophysical merger rates are too low to robustly establish a $\sigma_{\tilde{\Lambda}}$ versus $\tilde{\Lambda}_{\rm true}$ relationship \cite{LIGOScientific:2025slb}. As an alternative, we simulate $10{,}000$ representative BNS events using the uniform mass grid described in \secname~\ref{subsec:multid}. For each event, the EoS ensemble described in \secname~\ref{subsec:eos} is used to sample the corresponding $\Lambda_1$ and $\Lambda_2$ values. All signals are injected at a fiducial distance of $131\ \rm Mpc$, chosen as the maximum of the derivative of the detector efficiency curve for the HLV network considered here. Since most events are expected near this distance, we use $131\ \rm Mpc$ for our Fisher injections to generate a representative sample of SNRs, providing a meaningful estimate of the average behaviour of the detector. All other source parameters, apart from masses and tidal deformabilities, are kept fixed to enable a meaningful comparison.

We employ the Fisher matrix formalism \cite{Cutler:1994ys, Poisson:1995ef} to empirically derive the distribution of uncertainties in GW source parameters and SNRs for our assumed BNS population. While Bayesian inference is generally preferred for estimating uncertainties, the Fisher approach is computationally efficient and well-suited for the large number of events considered here.

\begin{figure}[t]
    \centering
    \includegraphics[width=\linewidth]{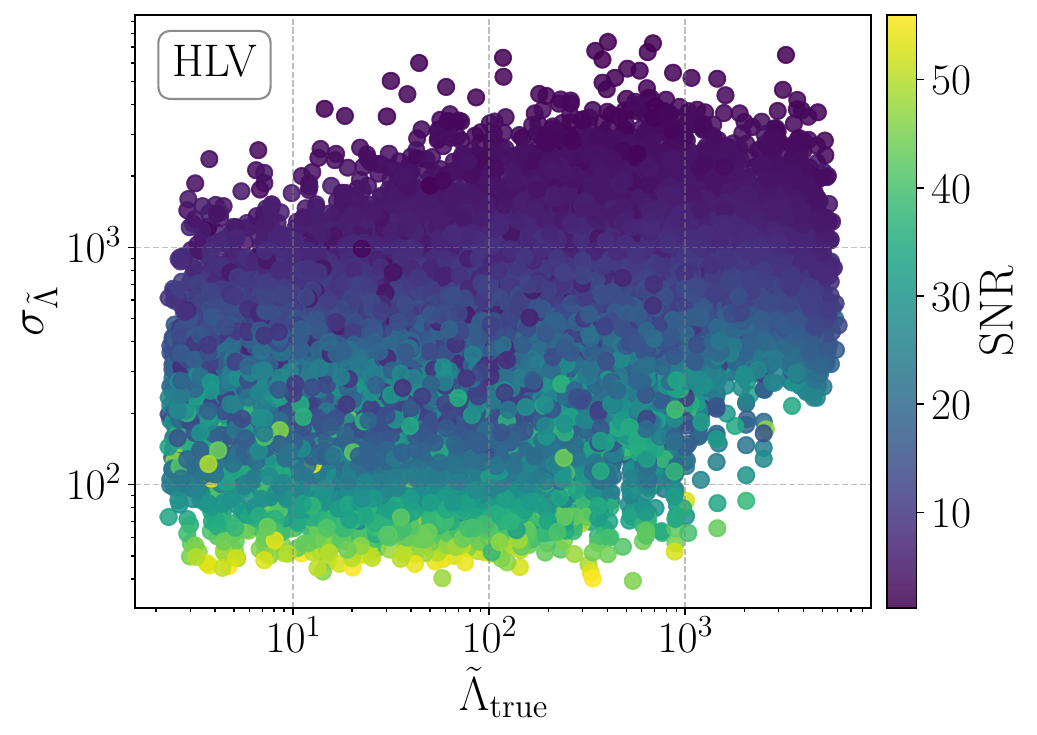}
    \caption{Variation of $\sigma_{\lt}$ with respect to the true $\lt_{\rm true}$ injected for the analysis. The colour variation depicts the SNRs of the considered signals. }
    \label{fig:app1}
\end{figure}

To ensure consistency, the \textsc{IMRPhenomXAS\_NRTidalv3} waveform model is used in the Fisher analysis. We evaluate $\sigma_{\tilde{\Lambda}}$ for the current-generation LVK detector network. The expected variation of $\sigma_{\tilde{\Lambda}}$ as a function of the injected $\tilde{\Lambda}$ is shown in \figname~\ref{fig:app1}.
\section{Origin of the Induced Prior on the Tidal Deformability}
\label{app2}

In this appendix, we provide a simple physical argument to explain the approximate form of the prior distribution induced on the tidal deformability parameters. This discussion is intended to provide intuition for the structure of the prior and does not replace the numerical prior adopted in the main analysis, which is derived directly from the sampling procedure and shown in \figname~\ref{fig:3_prior_comparison}.

The dimensionless tidal deformability $\Lambda$ of a NS of mass $m$ is given by Eq.~\ref{eq:lambda}. For a wide class of realistic EoSs, the NS radius $R$ varies only weakly with mass. Under this approximation, the dominant mass dependence of $\Lambda$ arises from the explicit $m^{-5}$ scaling, with a sub-leading contribution from the mass dependence of the quadrupolar tidal Love number $k_2$.

Approximating the Love number as a power law in mass, $k_2 \propto m^{-\alpha}$, the tidal deformability scales as
\begin{equation}
\Lambda \propto m^{-5-\alpha}.
\end{equation}
If the component masses are assigned a prior that is uniform, or only slowly varying, over the relevant mass range, this scaling induces an approximately power-law prior on $\Lambda$.

In GW analyses, the quantity directly constrained by the waveform is not the individual tidal deformabilities $\Lambda_i$, but the mass-weighted tidal parameter $\tilde{\Lambda}$, defined in Eq.~\ref{eq:lt}. This parameter is a linear combination of the individual $\Lambda_i$, weighted by powers of the component masses. Since each $\Lambda_i$ exhibits a similar power-law dependence on its corresponding mass, the resulting combination preserves the approximately power-law structure of the induced prior. As a result, the prior distribution of $\tilde{\Lambda}$ shown in \figname~\ref{fig:3_prior_comparison} falls off as a power law above its peak, the peak itself being set by the maximum NS mass in the ensemble and the detector-sensitivity broadening.

We emphasise that this derivation relies on simplifying assumptions regarding the mass dependence of the neutron star radius and Love number, and is intended solely as a qualitative explanation. In practice, the prior on $\tilde{\Lambda}$ used in this work is obtained numerically from the full parameterisation employed in the analysis.

\section{Handling detector and source frame parameters}
\label{app:detector_source_frame}

Using any EoS‑motivated or physically motivated prior that samples in source frame parameters requires repeated conversion between source and detector frame, specifically between luminosity distance \(D_L\) and redshift \(z\). In {\sc Bilby}, cosmological conversions such as \texttt{bilby.gw.conversion.luminosity\_distance\_to\\\_redshift} rely on {\sc Astropy}'s cosmology routines, which perform numerical integration for each call \citep{Astropy:2013muo}. When the prior is defined in source frame chirp mass rather than detector frame chirp mass, these conversions are invoked for every prior sample, including during normalisation factors such as the constraint factor calculations used in likelihood and prior consistency calculations. For a typical run with millions of samples, this results in tens of thousands of cosmology evaluations per sample, making sampling prohibitively slow without optimisation.

To alleviate this computational bottleneck, we use {\tt wcosmo} \citep{wcosmoPyPI}, a package developed for accelerated cosmological calculations compatible with {\sc Bilby} and related inference frameworks. {\tt wcosmo} implements efficient, grid‑based forward and inverse conversions between \(D_L\) and \(z\) that are orders of magnitude faster than repeated numerical integration, enabling fast interpolation of the luminosity distance–redshift relation. In practice, we initialize a tabulated grid spanning the full range of luminosity distances supported by our prior and use this grid for all conversions between \(D_L\) and \(z\) during sampling.

Within {\sc Bilby}, cosmological conversions are required both in the prior and in the construction of derived quantities, such as source frame masses and redshift from sampled detector frame parameters. Without the grid accelerations provided by {\tt wcosmo}, these conversions dominate the runtime when the prior samples in source frame, preventing efficient inference. By adopting {\tt wcosmo}, we retain compatibility with standard {\sc Bilby} workflows while significantly reducing the computational cost of cosmological conversions necessary for \priorname\ and other source‑frame priors.

\sk{This conversion is implemented for the full parameter space, including the extrinsic parameters of the 13- and 17-dimensional runs that will be sampled in the companion analysis of real events \citet{Khadkikar:2026san}. This implementation does not affect the analysis presented in this manuscript since the luminosity distance is fixed per injection to fix the SNR rather than sampled.}
\section{Issues with post-processing the uniform prior}
\label{app:reweighting}

\begin{figure*}[t]
    \centering
    \includegraphics[width=\linewidth]{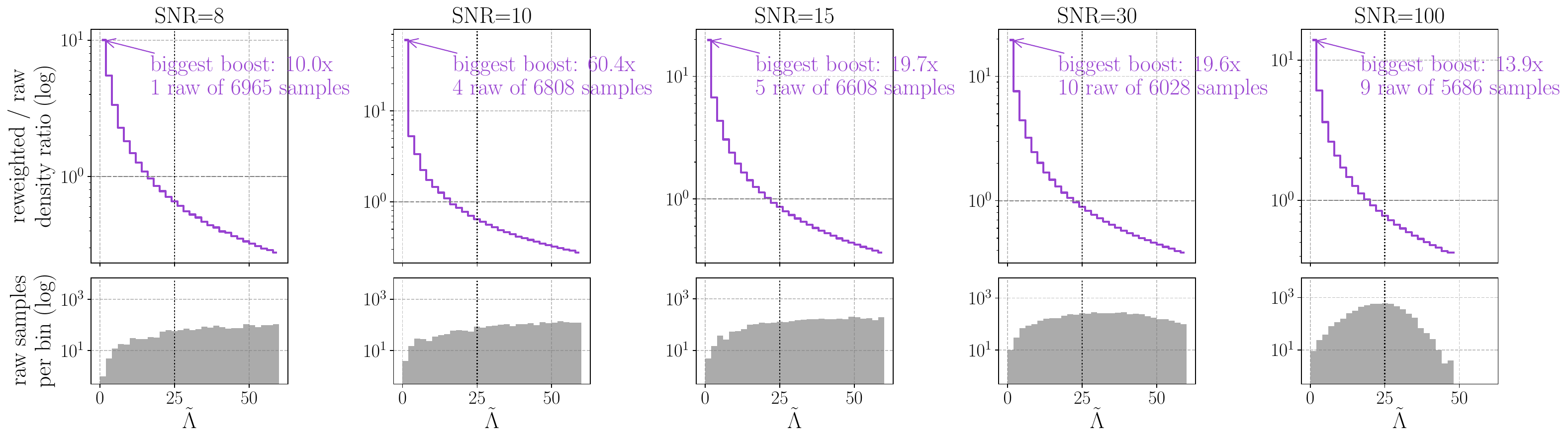}
    \caption{\sk{Diagnostics for the reweighted uniform posterior of a symmetric $1.4$--$1.4~\rm M_{\odot}$ system with $\lt_{\rm inj} = 25$, at network SNRs of 8, 10, 15, 30 and 100 (left to right). Lower panels show the number of raw samples per $\lt$ bin. Upper panels show the ratio of the reweighted to the raw density per bin, with the largest single-bin amplification annotated. The largest amplifications occur at low $\lt$, where the raw samples are fewest, reaching factors of 10, 60, 20, 20, and 14 from as few as one to ten samples across the five SNRs. The dotted vertical line marks $\lt_{\rm inj} = 25$.}}
    \label{fig:appendix_reweighting}
\end{figure*}

\sk{Throughout \secname~\ref{sec:results} we compare against a reweighted uniform posterior, constructed by importance-reweighting samples drawn under the implied uniform prior to mimic a true uniform prior. This appendix quantifies why that reweighted posterior is unreliable in the low-$\lt$ regime, supporting the argument made in \secname~\ref{subsec:reweighting}.}

\sk{Reweighting redistributes the samples an analysis has already drawn rather than generating new ones. Each sample is assigned a weight equal to the ratio of the target to the sampling prior, and the effective number of samples there sets the reliability of the reweighted density in any region. The implied uniform prior places little support at low $\lt$, so a uniform-prior analysis draws few samples in that region, and reweighting must reconstruct the low-$\lt$ posterior by applying large weights to a small number of points.}

\sk{\figname~\ref{fig:appendix_reweighting} makes this explicit. The lower panels show the number of raw samples per $\lt$ bin, which falls to single digits in the lowest bins at every SNR. The upper panels show the ratio of the reweighted to the raw density in each bin. This ratio is largest at low $\lt$, precisely where the samples are fewest. At SNR 8 the largest single-bin amplification is a factor of 10, applied to a bin containing one raw sample. At SNR 10 it reaches a factor of 60 from four samples, and comparable amplifications occur at the higher SNRs. A density reconstructed by amplifying one or a few samples by such factors reflects the sampling noise of those points rather than the underlying distribution, which is the origin of the spurious low-$\lt$ structure seen in the reweighted posteriors throughout \secname~\ref{sec:results}.}

\sk{The problem does not diminish with SNR. Even at SNR 100 the largest amplification is a factor of $14$ from nine samples, since the loud-signal posterior concentrates at low $\lt$ and the sparsely sampled bins remain sparsely sampled. Reweighting therefore cannot be relied upon to recover the low-$\lt$ region at any SNR considered here.}

\sk{This limitation could in principle be reduced by increasing the number of live points, so that the sampler populates the low-$\lt$ region more densely before reweighting. The analyses here already use $2048$ live points, matching the setting used in LVK compact binary production runs, and raising it further increases the computational cost and wall-clock time of every analysis. The \priorname\ provides correct support at low $\lt$ with the same $2048$ live points, without any post-processing, and we observe the fastest convergence and consistently accurate recovery among the priors compared. Achieving reliable low-$\lt$ inference is therefore better addressed through the prior itself than through reweighting a uniform-prior posterior.}

\begin{figure*}
    \centering
    \includegraphics[width=\linewidth]{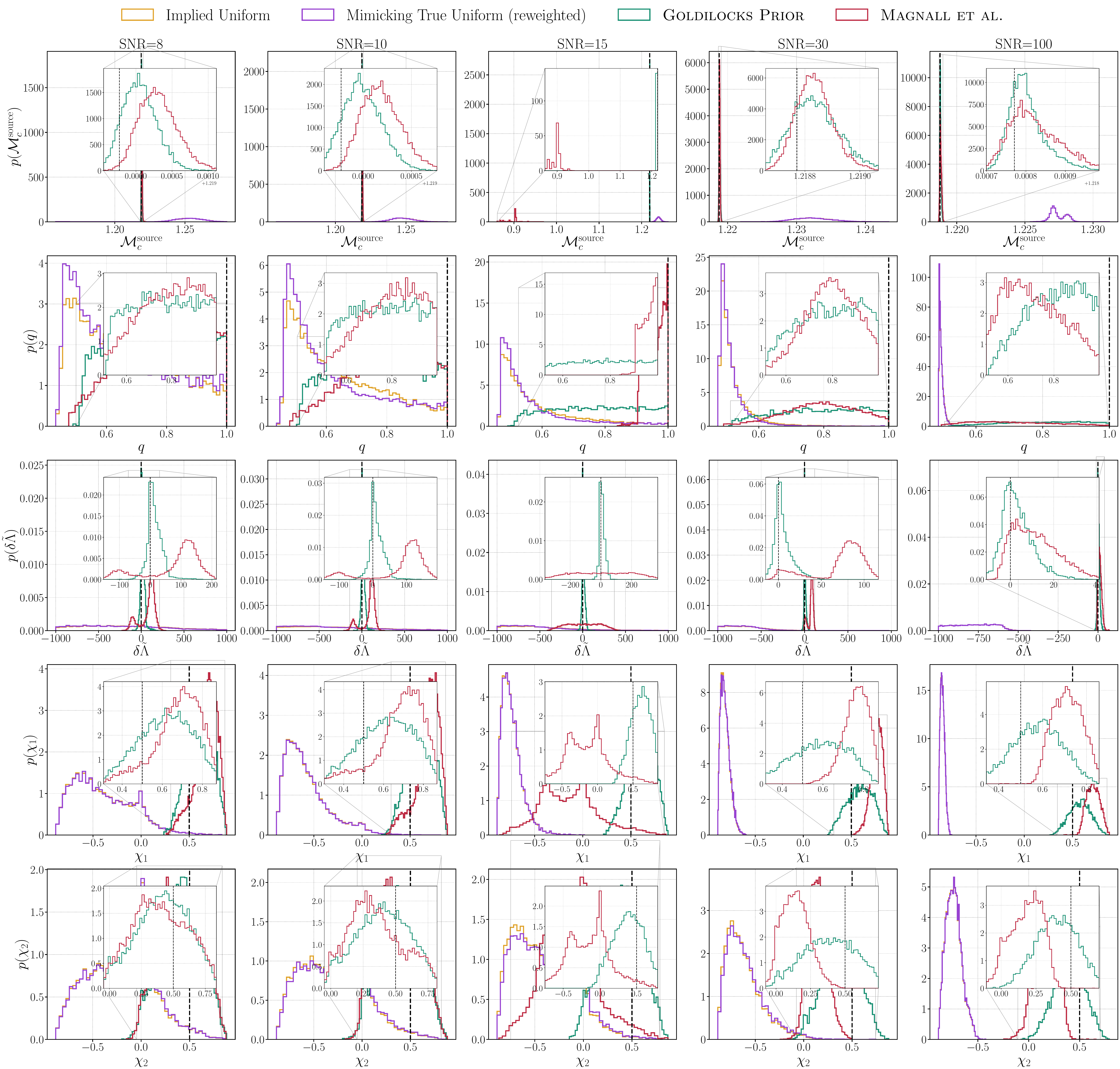}
    \caption{\sk{Recovered posteriors for the high-spin symmetric $1.4$--$1.4~\rm M_{\odot}$ system with $\chi_i = 0.5$ and $\lt_{\rm inj} = 25$, at network SNRs of 8, 10, 15, 30 and 100 (left to right). Rows show the source-frame chirp mass $\mc$, the mass ratio $q$, the reduced tidal deformability $\dlt$, and the two component spins $\chi_1$ and $\chi_2$. The recovered $\lt$ for this system is shown in \figname~\ref{fig:7_highspin}. Curves follow \figname~\ref{fig:4_lt25_symm}, and dashed vertical lines mark the injected values. Under the implied uniform and reweighted priors, every parameter is recovered incorrectly, and together, the $\mc^{\text{source}}$ is biased high, $q$ toward strong asymmetry, and both $\chi$s toward large negative values, the inverse correlations that reproduce a similarly short waveform. The prior of \citet{Magnall:2025zhm} recovers $\mc^{\text{source}}$ correctly but is offset in $\dlt$ and the spins. The \priorname\ is the only prior that recovers all parameters consistently with their injected values across the full SNR range.}}
    \label{fig:appendix_highspin}
\end{figure*}
\end{document}